\documentclass[letterpaper,twocolumn,10pt]{article}
\usepackage{usenix}
\usepackage{epsfig}
\usepackage{amsmath}
\usepackage{algpseudocode}
\usepackage{amssymb}
\usepackage{svg}
\usepackage{booktabs}
\usepackage{endnotes}
\usepackage{algorithm}
\usepackage{algpseudocode}     
\usepackage{algorithmicx}
\usepackage{amsmath}           
\usepackage{graphicx}          
\date{} 
\begin{document}

\title{\texttt{FastCache}: Optimizing Multimodal LLM Serving through Lightweight KV-Cache Compression Framework}

\author{
{\rm Jianian Zhu}\thanks{These authors contributed equally to this work.}$^{*}$\\
Huazhong University of Science and Technology
\and
{\rm Hang Wu}$^{*}$\\
Xidian University
\and
{\rm Haojie Wang}\\
Tsinghua University
\and
{\rm Yinghui Li}\\
Tsinghua University
\and
{\rm Biao Hou}\\
Xidian University
\and
{\rm Ruixuan Li}\\
Huazhong University of Science and Technology
\and
{\rm Jidong Zhai}\\
Tsinghua University
}

\maketitle

\thispagestyle{empty}

\section*{Abstract}
Multi-modal Large Language Models (MLLMs) serving systems commonly employ KV-cache compression to reduce memory footprint. However, existing compression methods introduce significant processing overhead and queuing delays, particularly in concurrent serving scenarios. We present \texttt{FastCache}, a novel serving framework that effectively addresses these challenges through two key innovations: (1) a dynamic batching strategy that optimizes request scheduling across prefill, compression, and decode stages, and (2) an efficient KV-cache memory pool mechanism that eliminates memory fragmentation while maintaining high GPU utilization. Our comprehensive experiments on the GQA and MileBench datasets demonstrate that \texttt{FastCache} achieves up to 19.3$\times$ reduction in Time-To-First-Token (TTFT) and 12.1$\times$ improvement in throughput compared to state-of-the-art baselines. The system maintains stable performance under high-concurrency scenarios (up to 40 req/s) while reducing average memory consumption by 20\%. These results establish \texttt{FastCache} as an efficient solution for real-world LLM serving systems with KV-cache compression.

\section{Introduction}
\label{intro}

Modern Multimodal Large Language Models (MLLMs) have achieved impressive performance on various generation~\cite{liu2023llava, openai2023gpt} and multimodal comprehension tasks.
However, as these models become larger and more versatile, they typically maintain a sizable \emph{KV-cache} (key-value cache)~\cite{pope2023efficiently} during inference.
This cache usage grows quickly with both the length of the input sequence and the number of concurrent requests, posing significant memory bottlenecks in practical deployment~\cite{kwon2023efficient, zheng2024sglang, patel2024splitwise, zhong2024distserve, lee2024infinigen}.
In scenarios of real-time \emph{high concurrency}, the KV-cache can rapidly inflate to the point of straining GPU or system memory~\cite{kwon2023efficient}.

Existing solutions for reducing KV-cache memory footprint can be categorized into two main approaches: memory pooling/sharing and KV-cache compression. 
Memory pooling approaches, as implemented in systems like vLLM and Mooncake, primarily focus on reducing memory fragmentation through efficient memory management and request scheduling. These methods only optimize memory allocation patterns without actually reducing the underlying KV-cache data size, limiting their effectiveness under intensive memory pressure from concurrent requests.

\begin{figure}[ht]
\centering
\includegraphics[width=0.75\linewidth]{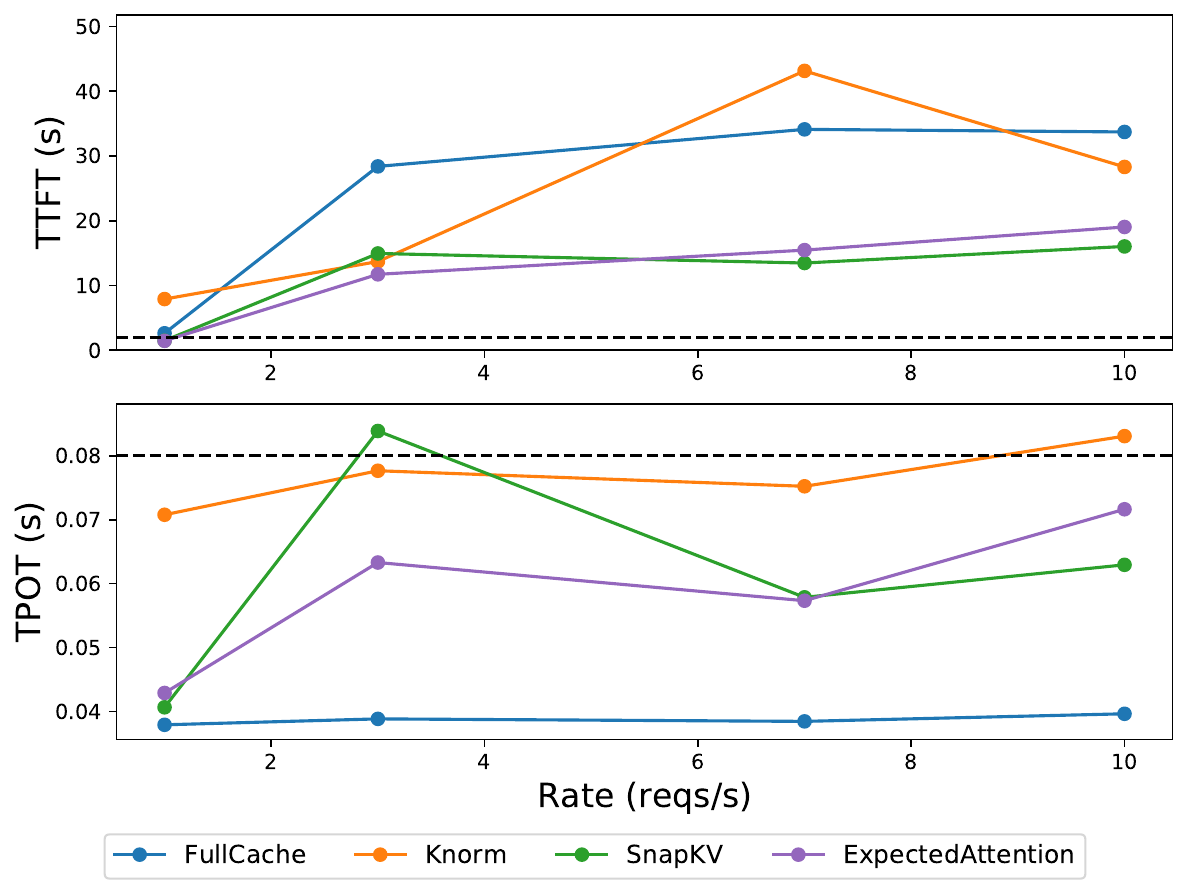}
\caption{Performance comparison of the state-of-the-art KV-cache compress methods~\cite{li2024snapkv, devoto2024simpleeffective, kvpress2024} for LLaVA-1.5-7B under real-time serving system.}
\label{fig:init}
\end{figure}

Recent KV-cache compression methods typically achieve compression ratios of 2-4$\times$ through various techniques. For example, common approaches include reducing precision from float32/float16 to int8 quantization with minimal accuracy loss, and selectively pruning less important KV-cache entries based on attention patterns. While KV-cache compression methods can effectively reduce the memory footprint to 25-50\% or even lower of the original size~\cite{li2024snapkv}, they suffer from \textbf{striking Time-To-First-Token (TTFT) problem}, which obstructs their deployments in product environment. KV-cache compression introduces an additional compression step, which obviously prolongs the end-to-end interference time. This leads to an augmentation of the average TTFT , significantly deteriorating the user experience. 

To quantify this performance bottleneck in MLLM serving scenarios, we conducted the experiment to evaluate the average TTFT and TPOT performance of state-of-the-art KV-cache compression approaches (Knorm~\cite{devoto2024simpleeffective}, SnapKV~\cite{li2024snapkv}, and ExpectedAttention~\cite{devoto2024simpleeffective}) using the LLaVA-1.5-7B model on a single NVIDIA H100 GPU. As shown in Figure~\ref{fig:init}, the TTFT results reveal significant performance variations across different methods. At high request rates (6-8 req/s), Knorm's TTFT peaks at approximately 42 seconds, while FullCache maintains a more stable but still high TTFT around 33 seconds. ExpectedAttention demonstrates better performance with TTFT staying below 20 seconds across all request rates. However, all methods still substantially exceed the typical Service Level Objective (SLO) of 2 seconds, indicated by the dashed line in the figure. The TPOT measurements show similar variations, with FullCache maintaining the lowest and most stable TPOT around 0.04 seconds, while other approaches fluctuate between 0.06-0.08 seconds as request rates increase.

Through systematic analysis of these experimental results and thorough examination of existing approaches, we identify that existing KV-cache compression approaches suffer from long TTFT problem due to the following two fundamental drawbacks:

\textbf{(1) Attention-centric compression limitations:} Existing approaches heavily rely on attention patterns for compression decisions. This leads to sequential processing of attention maps for each request, creating inherent bottlenecks under high concurrency scenarios, making them unsuitable for on the fly or real-time serving systems.

\textbf{(2) Inefficient compression scheduling:} Current approaches typically follow a first-come-first-served (FCFS) compression strategy, triggering individual compression operations for each request. This not only incurs repeated compression overhead but also fails to leverage potential opportunities for batch compression optimization, making them inefficient for online serving under high concurrency.


To fill this gap, we propose \emph{resource-aware KV-cache compression framework} that offers an \emph{efficient} management of the KV-caches for MLLM. Our framework embodies the following key features: 
\begin{itemize}
    \item \textbf{Linear Compression,} we introduce a novel compression algorithm that does not rely on attention score analysis. By rethinking the compression criteria, our method enables batching processing of KV-cache entries, eliminating the sequential bottlenecks in existing approaches. This allows for faster compression times and better scalability in high-concurrency environments.
    \item \textbf{Resource Aware KV-cache Memory Pool,} our framework incorporates an resource-aware pipeline that dynamically adjusts KV-cache memory pool based on real-time system resource metrics. 
\end{itemize}

By leveraging online scheduling and parallel execution, our framework minimizes compression overhead and maximize throughput. This ensures low latency and maintains high throughput, meeting the SLO requirements even under heavy concurrent loads.


We implement the resource-aware KV-cache compression framework and conducts extensive evaluations on single-producer single-consumer queue (SPSC)~\cite{mitropoulou2016lynx}. The evaluation results show that our system significantly outperforms state-of-the-art methods across all key metrics: \textbf{achieving 12.1$\times$ higher throughput}, \textbf{reducing TPOT by more than 2.2$\times$}, and \textbf{decreasing TTFT by more than 19.3$\times$} compared to the best performing baseline under static batching configurations. These substantial improvements demonstrate the effectiveness of our approach in addressing the bottlenecks of existing rigid KV-cache compression system.

The remainder of this paper is organized as follows: We first introduce the background and motivations (\S~\ref{Back}). 
Building on these insights, we propose our framework (\S~\ref{design}), demonstrating how it effectively addresses the performance bottlenecks in concurrent serving scenarios while maintaining high compression quality. We evaluate our framework extensively (\S~\ref{imple} and \S~\ref{experi}) and conclude with discussions (\S~\ref{related} and \S~\ref{conclu}).

\section{Background and Motivation}\label{Back}
\subsection{Background}
\begin{figure}[ht]
    \centering
    \includegraphics[width=0.5\textwidth]{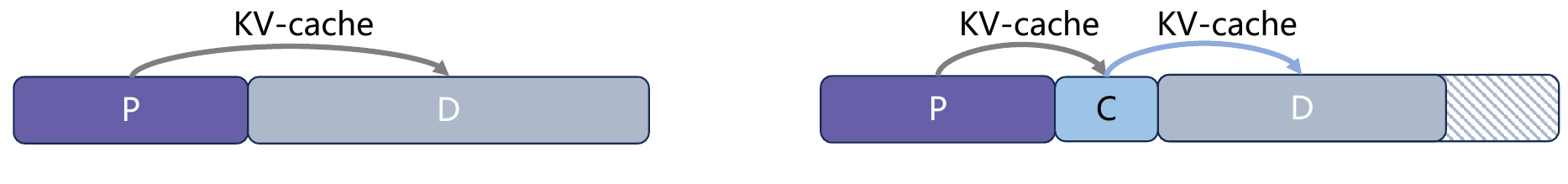}
    \caption{Illustration of KV-cache transmission in different serving scenarios. Left: Traditional serving pipeline with prefill (P) and decode (D) stages; Right: Serving pipeline with compression, where a compression stage (C) is introduced between prefill and decode stages, requiring careful design to balance memory savings and computational overhead.}
    \label{fig:problems}
\end{figure}
\textbf{KV-cache Memory Footprint Analysis.} The accumulation of KV-cache during inference poses significant challenges for high-concurrency scenarios. The memory footprint of the KV-cache grows linearly with both the sequence length and the model size. Without proper compression and management strategies, the KV-cache can quickly exhaust GPU memory, leading to degraded service quality or system failures. For instance, handling 1M tokens with Llama 3.1-70B requires up to 330GB of memory for KV-cache~\footnote{https://github.com/NVIDIA/kvpress/tree/main}, highlighting the severe memory constraints in long-context scenarios. 

The characteristics of KV-cache access patterns influence the system design. During generation, the model frequently reads from the cache but only writes new entries sequentially, creating an asymmetric access pattern. As shown in Figure~\ref{fig:problems}, introducing a compression stage (C) between prefill (P) and decoding (D) can reduce KV-cache memory footprint and potentially decrease decoding time. However, if the compression method is not carefully designed, the additional computational overhead might outweigh the benefits gained from reduced memory usage, leading to increased overall latency instead of the intended performance improvement.

\begin{figure}[ht]
    \centering
    \includegraphics[width=0.5\textwidth]{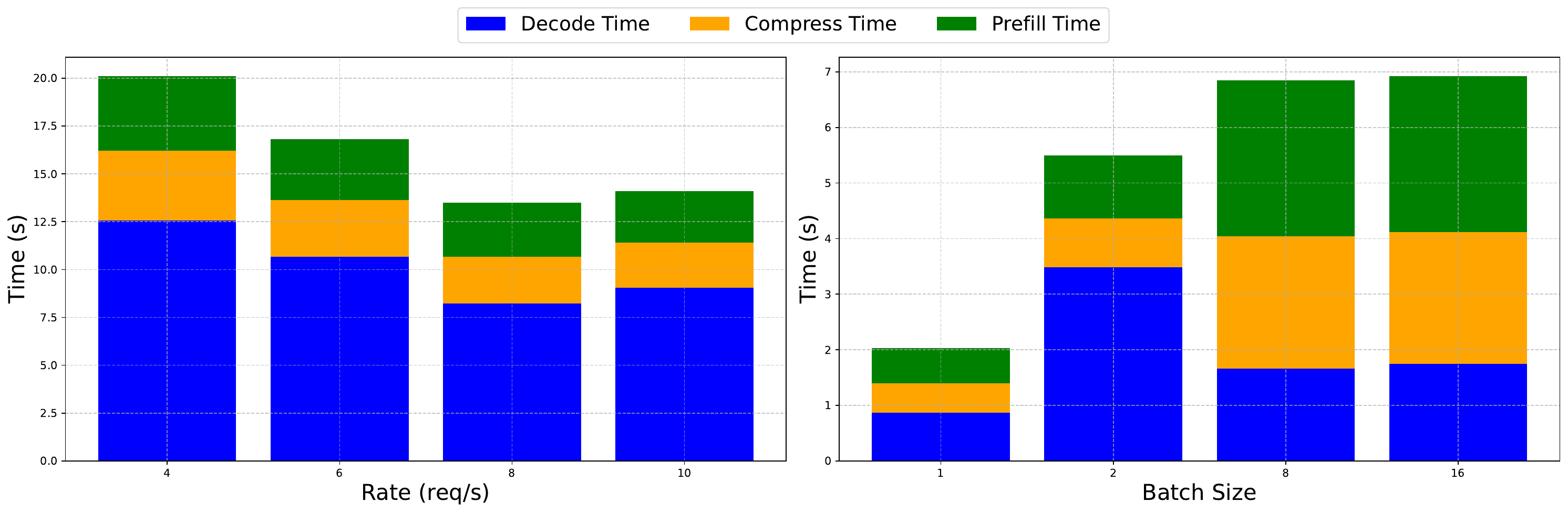}
    \caption{Processing time breakdown showing three stages (decode, compression, and prefill) under different scenarios. Left: Time distribution with varying request rates (4-10 req/s); Right: Time distribution across different batch sizes (1-16), where the maximum \texttt{input\_length} = 128, and we chose the advanced kv-compress method~\cite{kvpress2024}.}
    \label{fig:batch_breakdown}
\end{figure}

\textbf{KV-Cache Compression Overhead Analysis.}
\label{2.2}
The predominant KV-cache compression techniques employ attention-based mechanisms.
These methods operate by computing attention scores to evaluate the importance of key-value pairs.
The computational complexity of this approach scales with $O(L × N × H × dh)$ per request, where $L$ represents sequence length, $N$ denotes batch size, $H$ indicates the number of attention heads, and $dh$ represents the dimension per head. This substantial computational overhead manifests in two critical ways: first, the processing cost of computing attention scores 
for each token, and second, the resource contention between compression operations and core inference tasks.
Our empirical analysis reveals the practical implications of this overhead in real-world serving scenarios. 
As demonstrated in Figure~\ref{fig:batch_breakdown}(left), 
the compression stage adds 3.662 seconds to the total processing latency at 4 requests per second, significantly degrades the latency performance. 
The compression process creates a secondary computational bottleneck alongside the primary inference task, competing for the same GPU resources and leading to suboptimal hardware utilization.
This performance impact becomes particularly pronounced in service where maintaining consistent latency and meeting SLO is crucial. While KV-cache compression is essential for managing memory in long-sequence and massive request scenarios, current attention-based approaches introduce a significant performance degradation. Such findings underscore the urgent need for more efficient compression frameworks. 

\begin{figure}[ht]
\centering
\includegraphics[width=0.5\textwidth]{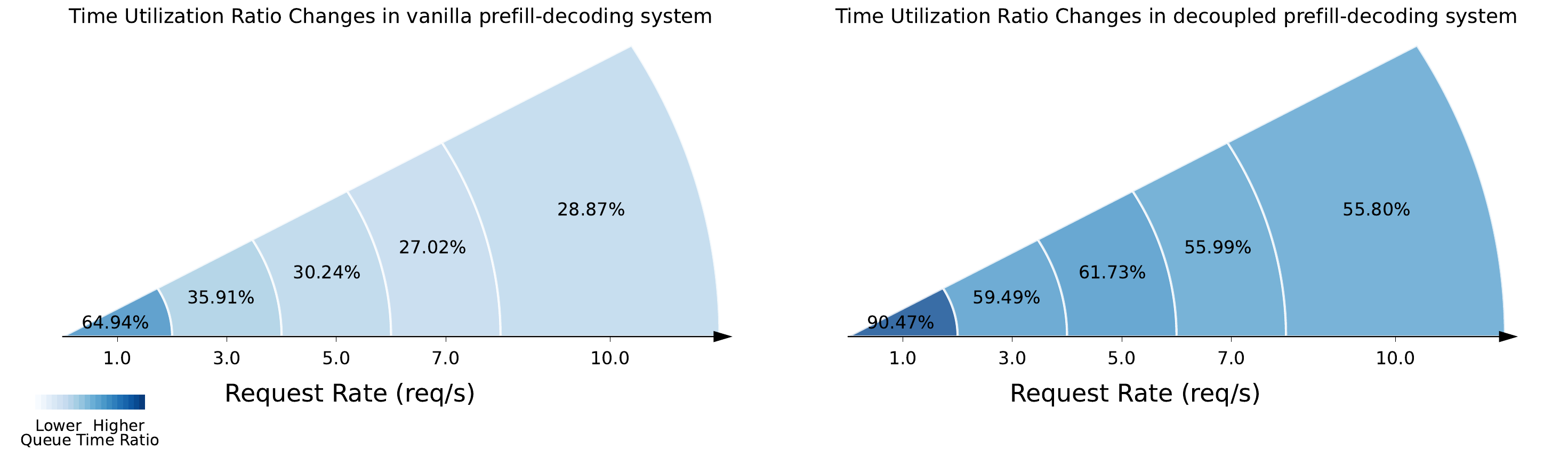}
\caption{Comparison of effective time utilization ratios between vanilla sequential processing (left) and decoupled prefill-decoding (right) systems across different request rates with KV-cache compression enabled.}
\label{fig:time_utilization}
\end{figure}

\subsection{Analysis of System Bottlenecks}\label{problem}
Current KV-cache compression based LLM serving systems face significant challenges in concurrency scenarios due to two fundamental limitations: inefficient pipeline processing and rigid memory management. Through comprehensive evaluation of serving system performance, we have identified that existing approaches fail to provide stable processing times and frequently result in system underutilization under realistic serving conditions. As shown in Figure~\ref{fig:time_utilization}, this inefficiency is particularly evident in vanilla sequential processing systems where prefill and decoding stages are tightly coupled. In such systems, the effective time utilization ratio remains notably low, reaching only 28.87\% at 10 req/s, indicating substantial time spent in request queuing rather than actual processing. In contrast, when prefill and decoding stages are decoupled, the system achieves significantly higher utilization ratios, reaching 55.80\% at the same request rate. This stark difference in utilization ratios demonstrates that the coupling of prefill and decoding stages in conventional systems creates a bottleneck that severely impacts system efficiency, particularly when KV-cache compression is enabled. The increased queuing time in the coupled system suggests that compression operations further exacerbate the pipeline inefficiencies, while a decoupled architecture better manages these overheads by allowing for more flexible resource allocation and reduced contention.

\begin{figure}[ht]
    \centering
    \includegraphics[width=0.35\textwidth]{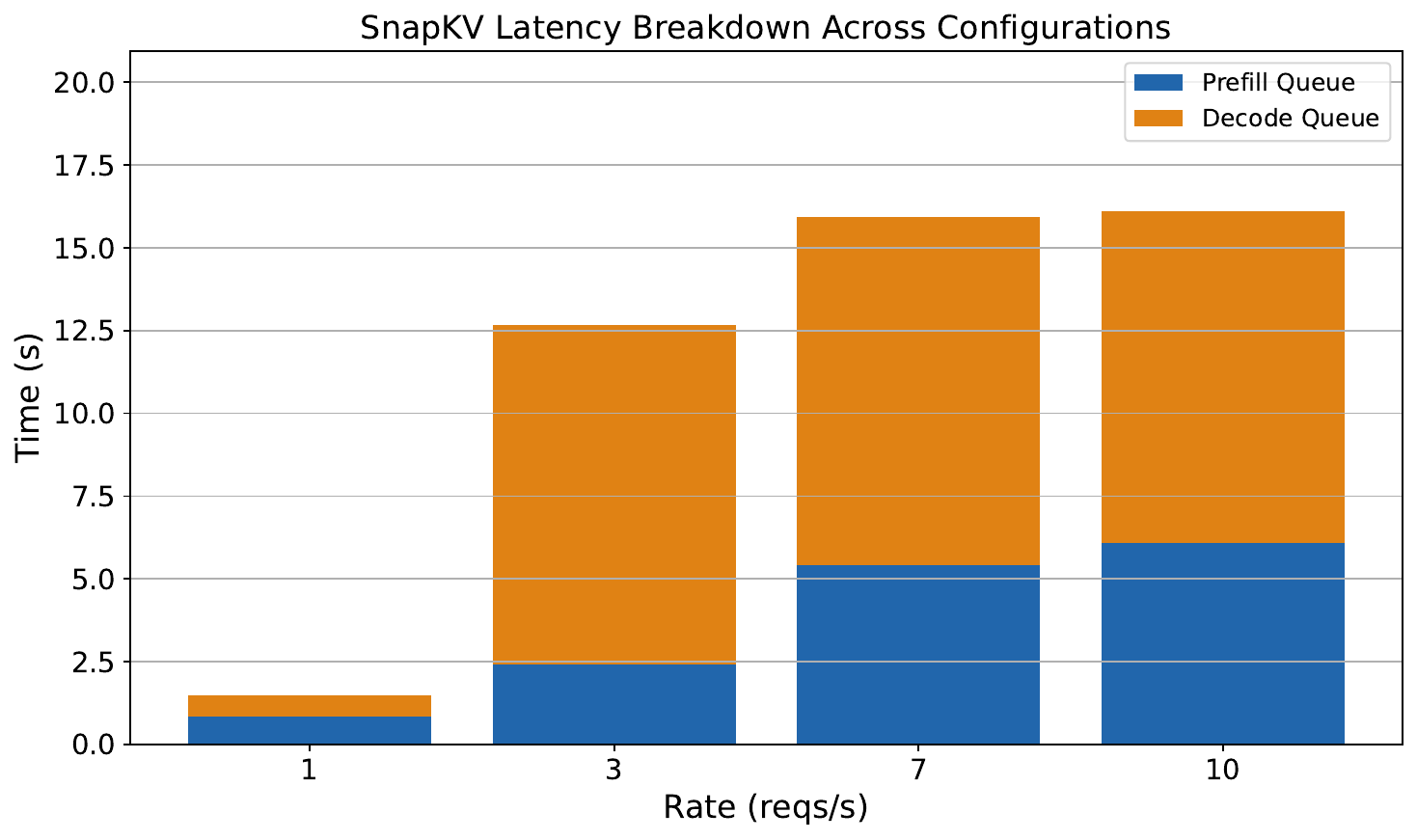}
    \caption{Latency breakdown for the SnapKV compression method within a prefill-decode separated architecture. 
    }
    \label{fig:queue_analysis}
\end{figure}

\subsubsection{Processing Pipeline Bottlenecks}
\label{over}

However, merely decoupling prefill and decode stages does not fully address the system's efficiency challenges. As illustrated in Figure~\ref{fig:queue_analysis}, significant queuing delays persist within the decoupled architecture. Our detailed analysis reveals that approximately 40\% of the total request latency is still attributable to queuing rather than actual processing time, with these delays distributed across both prefill and decode stages.
Even state-of-the-art compression methods like SnapKV~\cite{li2024snapkv}, despite operating within a prefill-decode separated architecture, fail to fully utilize available computational resources. This underutilization occurs primarily due to two factors: First, the rigid First-Come-First-Serve (FCFS) scheduling model creates artificial bottlenecks, forcing requests to wait sequentially even when parallel processing capacity is available. Second, delays in the prefill stage create a cascading effect, propagating through the compression phase and ultimately impacting decode performance. This suggests that while architectural decoupling is beneficial, it must be complemented by more sophisticated scheduling and resource management strategies to achieve optimal system performance.

\subsubsection{Memory Management Challenges}
\label{inf}

The sequential nature of current LLM serving pipelines leads to three critical resource management challenges:

\textbf{Suboptimal Resource Utilization.}
Without the ability to interleave different processing stages across multiple requests, the system fails to maximize GPU utilization. This inefficiency is particularly pronounced during operations like prefill and compression, where the sequential processing model forces the system to handle one request at a time. 

\textbf{Compression Performance Overhead.}
The integration of compression operations into the LLM serving pipeline introduces significant performance overhead and stability challenges. 
Existing KV-cache compression approaches cannot maintain consistent processing times, making them unsuitable for real-time serving systems.

\textbf{Inefficient Memory Management.} 
The simultaneous retention of both pre-compression and post-compression cache states in VRAM outdated pre-compression caches, which we term "zombie KV-caches," remain in memory even after compression completes, effectively doubling the memory footprint per request. 
The lack of automatic mechanisms to identify and reclaim these zombie caches prevents the system from fully capitalizing on the benefits of compression.

\section{System Design}\label{design}
\begin{figure}[t]
   \centering
   \includegraphics[width=0.45\textwidth]{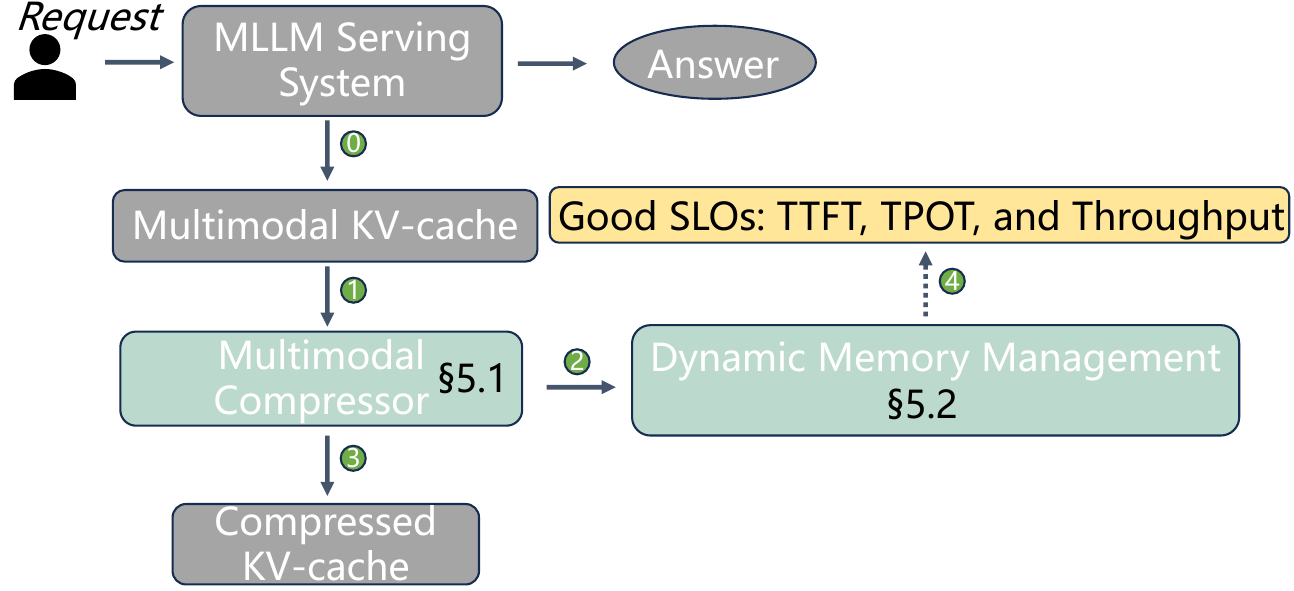}
   \caption{Overview of \texttt{FastCache}. The system processes multimodal requests through a lightweight compressor and dynamic memory manager to achieve optimal SLOs (TTFT, TPOT, and throughput). Numbers (0-4) indicate the processing flow: initial KV-cache generation (0), compression (1-3), and memory management for performance optimization (4).}
   \label{fig:overview}
\end{figure}

To address the achallenges, particularly the pipeline bottlenecks, memory redundancy, and inefficient resource management, we propose \texttt{FastCache}, an efficient KV-cache compression framework for multimodal LLM serving. As illustrated in Figure~\ref{fig:overview}, our system consists of two key components: a lightweight multimodal compressor (\S~\ref{5.1}) and a dynamic memory management module (\S~\ref{5.2}).

Our multimodal compressor efficiently handles KV-cache compression while preserving modality-specific features, addressing the high overhead challenge from previous methods. The dynamic memory management module tackles the memory redundancy problem by automatically tracking and reclaiming zombie KV-caches, while optimizing resource allocation to maximize concurrent request processing. Together, these components enable \texttt{FastCache} to maintain consistent performance metrics (TTFT, TPOT, and throughput) even under high-concurrency scenarios.

\subsection{Multimodal Compressor}\label{5.1}
\paragraph{Modality-Specific Design.}Our novel multimodal KV-cache compression approach is designed to effectively compress both text and image attention patterns while preserving their distinct characteristics. The core innovation lies in decoupling the compression process for different modalities, allowing the system to learn modality-specific compression patterns through separate attention-based MLPs.

For a given input sequence containing both image and text tokens, our method first separates the KV-cache based on modality boundaries. The image and text portions are then compressed independently using their respective compression networks. Specifically, for a compression factor of $k$, the method reduces sequences of length $n$ to length $n/k$ by reshaping and processing chunks of $k$ consecutive tokens. The compressed representations are finally concatenated to maintain the original sequence ordering, enabling seamless integration with the model's attention mechanism.

\paragraph{Offline Training.}
To achieve on the fly compression performance, we adopt a self-supervised learning approach~\cite{he2022masked}. Two specialized MLPs are trained independently - one for text and one for image modalities - using a masked prediction loss function. During training, portions of the input sequences are randomly masked, and the MLPs learn to reconstruct the masked tokens from their compressed representations, effectively learning modality-specific compression patterns. This self-supervised strategy enables the networks to discover efficient compression schemes without requiring explicit supervision. Once trained on a diverse set of multimodal data, these networks can perform compression on the fly without additional fine-tuning, making them suitable for real-time inference scenarios. This MLP-based compression approach is compute-intensive rather than memory-intensive, making it particularly well-suited for modern GPU architectures where computational throughput is abundant but memory bandwidth is often the bottleneck. The ability to trade memory access for computation allows our method to achieve faster compression speeds compared to traditional memory-bound compression algorithms.
\subsection{Dynamic Resource Management}\label{5.2}

\paragraph{Adaptive Strategy.}
Our system implements proactive GPU memory management and dynamic request scheduling to maximize resource utilization. Unlike traditional first-come-first-serve (FCFS) approaches that process requests sequentially or static batching methods that use fixed batch sizes~\cite{zhong2024distserve}, our system continuously monitors GPU memory availability and dynamically constructs optimal batch sizes based on real-time resource conditions.

\begin{figure}[t]
   \centering
   \includegraphics[width=0.45\textwidth]{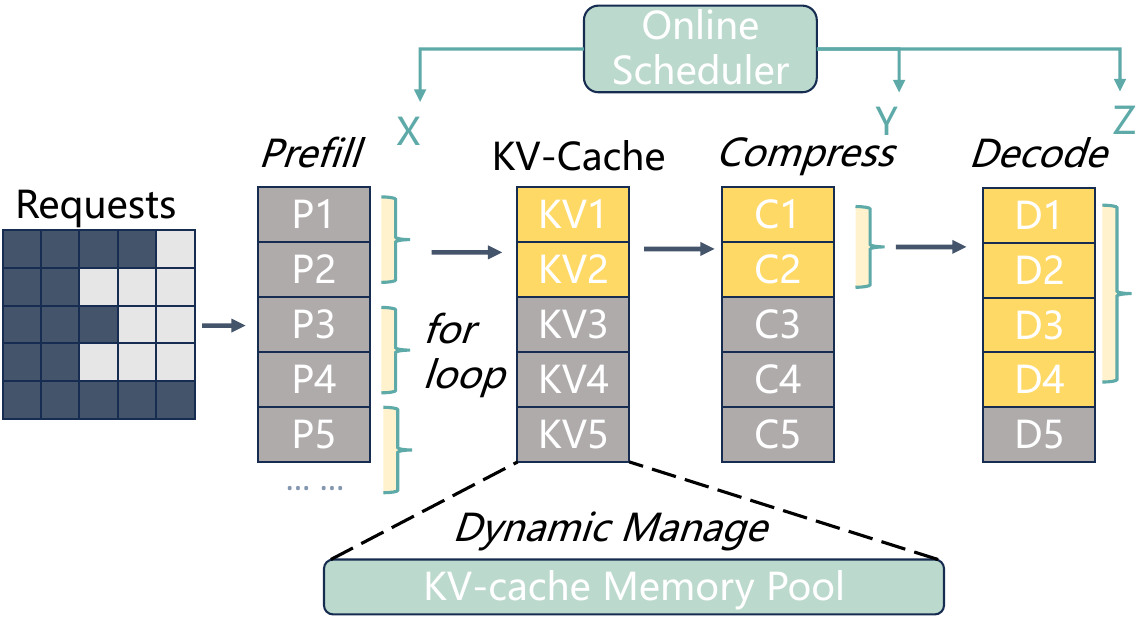}
   \caption{\texttt{FastCache} of dynamic scheduling pipeline. The \textit{Prefill}, \textit{Compress}, and \textit{Decode} queues are dynamically batched with sizes $X$, $Y$, and $Z$ determined by the online scheduler based on available resources. During \textit{Prefill}, requests are processed sequentially in a for-loop to generate KV-caches, which are then merged to match prefill batch size $X$. The \textit{Dynamic Manage} component, implemented through a KV-cache memory pool, prevents memory redundancy by automatically managing all KV-caches throughout their lifecycle, while the scheduler maintains optimal batch sizes for \textit{Compress} ($Y$) and \textit{Decode} ($Z$) stages.}
   \label{fig:pipeline}
\end{figure}

\paragraph{KV-cache Memory Pool.}
To address the memory redundancy problem where both compressed and uncompressed KV-caches coexist, we introduce a KV-cache memory pool that efficiently manages cache lifecycle. This pool serves two critical functions: automatically reclaiming zombie KV-caches (pre-compression caches) immediately after compression completes, and providing centralized management of compressed caches. As illustrated in Figure~\ref{fig:pipeline}, the memory pool dynamically manages all KV-caches throughout the pipeline stages, ensuring that only necessary caches remain in GPU memory. This approach effectively prevents memory waste from lingering pre-compression caches while maintaining optimal memory utilization for active requests.

\paragraph{Memory-Aware Scheduling.}
The memory monitoring component maintains a precise view of GPU memory usage by tracking active KV-caches through the memory pool, along with model parameters and intermediate computations. Based on this real-time memory map, the system calculates the maximum possible batch size that can be safely processed given current memory constraints. This adaptive approach ensures we fully utilize available GPU memory while preventing out-of-memory errors that plague static batching strategies.

Our online scheduling strategy maintains a dynamic request queue and implements a greedy batching mechanism that prioritizes memory utilization over request order. As depicted in Figure~\ref{fig:pipeline}, the scheduler orchestrates three key stages: prefill (batch size $X$), which processes requests sequentially to generate KV-caches, compress (batch size $Y$), which processes merged KV-caches, and decode (batch size $Z$). The memory pool actively manages KV-caches across all these stages, ensuring efficient memory utilization throughout the pipeline. When GPU memory becomes available, rather than immediately processing the next request in queue as in FCFS systems, our scheduler examines all pending requests and constructs the largest possible batch that fits within current memory constraints. This approach provides several advantages over FCFS scheduling: (1) higher throughput by maximizing parallel request processing, (2) better resource utilization by adapting batch sizes to available memory, and (3) reduced average latency by eliminating the head-of-line blocking common in FCFS systems. Through tight integration with the KV-cache memory pool, our system achieves efficient memory management while maintaining high throughput and low latency in multi-request scenarios.

\begin{figure}[t]
   \centering
   \includegraphics[width=0.45\textwidth]{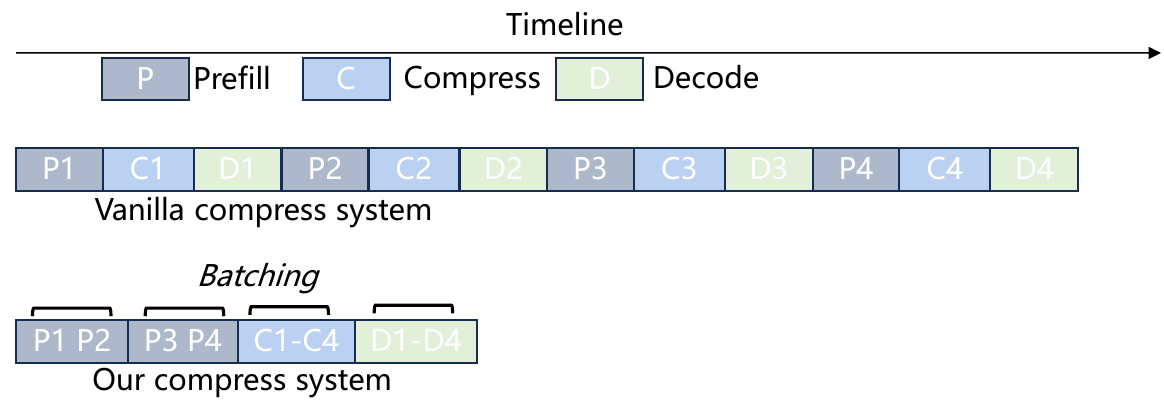}
   \caption{Comparison between vanilla and our compress system pipeline. Our system employs batching to group same operations together, reducing scheduling overhead.}
   \label{fig:timeline}
\end{figure}

\subsection{Pipeline Optimization} \label{pipe}
Traditional compression systems process requests through three sequential stages: prefill, compress, and decode. However, this vanilla pipeline introduces significant scheduling overhead and inefficient resource utilization due to sequential processing. As shown in Figure~\ref{fig:timeline}, the vanilla system handles each request independently in strict sequence (P1->C1->D1->P2->C2->D2...), leading to suboptimal GPU utilization and increased end-to-end latency.
Our system addresses these inefficiencies through comprehensive pipeline optimization with intelligent batching strategies. By leveraging our multimodal compression approach, we group similar operations together to maximize parallel execution potential. Rather than processing requests one at a time, our system employs a staged batching approach: first batching multiple prefill operations (P1-P2, P3-P4), then performing compressed operations in larger groups (C1-C4), and finally executing decode operations together (D1-D4).
This orchestrated batching strategy brings several key benefits. First, by processing similar operations together, we reduce the overhead of switching between different operation types. Second, batching enables more efficient use of GPU resources by increasing the effective batch size for each operation type. Finally, our approach reduces the total number of kernel launches and scheduling operations required to process the same number of requests. Through this integration of efficient operation batching and optimized pipeline execution, we achieve significant reductions in end-to-end latency while maintaining high throughput. The experimental results (\S \ref{experi}) show that our batched execution strategy substantially outperforms the traditional sequential approach in both resource utilization and overall processing efficiency. 
\begin{algorithm}[H]
\caption{\texttt{FastCache} for MLLM Serving}
\begin{algorithmic}[1]
\Require{Request stream $R$, Model $M$, Compressor $C$, Batch size bounds $[B_{min}, B_{max}]$, Compression ratio $\lambda$}
\Ensure{Generated responses with optimized latency-throughput}
\Function{ServeRequests}{$R$, $M$, $C$}
    \State Initialize KVCachePool $P$, Scheduler $S$
    \While{serving requests}
        \State $Q \gets$ \textsc{AccumulateRequests}($R$)
        \If{$|Q| \geq B_{min}$ \textbf{or} \textsc{WaitTimeExceeded}()}
            \State $batch \gets$ \textsc{FormBatch}($Q$, $B_{max}$)
            \State $kv \gets P.\textsc{Store}(M.\textsc{Prefill}(batch))$
            \State $kv_c \gets C(kv, \lambda)$ \Comment{Batch compression}
            \State $responses \gets$ \textsc{ExecuteDecoding}($M$, $kv_c$, $batch$)
            \State $P.\textsc{Release}(kv_c)$
        \EndIf
        \State $S.\textsc{AdjustBatch}()$ \Comment{Dynamic scheduling}
    \EndWhile
\EndFunction
\Statex
\Function{ExecuteDecoding}{$M$, $kv_c$, $batch$}
    \While{not complete}
        \State $tokens \gets M.\textsc{Generate}(batch, kv_c)$
        \State \textsc{UpdateKVCache}($kv_c$, $tokens$)
    \EndWhile
    \State \Return $tokens$
\EndFunction
\end{algorithmic}
\label{alg:kv-compression}
\end{algorithm}
To systematically realize these optimizations, we design a dynamic multi-stage batching algorithm that coordinates request processing across the pipeline stages (Algorithm~\ref{alg:kv-compression}). The algorithm maintains a KV-cache pool for efficient memory management where compressed KV-caches are stored and retrieved in batches. For incoming requests, instead of immediate processing, the system accumulates them until reaching an optimal batch size or maximum wait time threshold. This batched prefill operation generates initial KV-caches which are then compressed together to minimize compression overhead. During decoding, our dynamic scheduler adaptively adjusts batch sizes based on runtime conditions, balancing the tradeoff between processing efficiency and response latency. This coordinated approach enables efficient management of GPU memory while maximizing the benefits of batched operations across all pipeline stages.

\section{Implementation and Evaluation}\label{imple}

\subsection{System Implementation}
Our system implements a SPSC concurrent serving framework for commonly used MLLM models with dynamic batching and KV-cache compression in 6.5K lines of Python code. The system consists of a service middleware, execution engine, compression pipeline, and resource manager.

The service middleware implements online scheduling through a priority-based queue system. It monitors GPU memory availability to construct optimal batch sizes and maintains a memory map of active KV-caches, model parameters, and computations to prevent out-of-memory.

The execution engine employs dynamic memory management through a specialized KV-cache memory pool. Drawing inspiration from~\textit{memory pool} designs~\cite{li2023pond, cho2024starnuma}, we implement a centralized cache management system that automatically handles the lifecycle of KV-caches. This pool actively manages memory allocation and deallocation, particularly focusing on the efficient reclamation of zombie KV-caches post-compression while maintaining optimal memory utilization for active requests.

For compression, we leverage our proposed lightweight multimodal KV-cache compressor, which achieves a 5$\times$ compression ratio while maintaining model performance. The compressor effectively reduces the memory footprint of KV-caches to one-fifth of their original KV-cachge size without compromising generation quality.

Our greedy batching mechanism constructs maximum-sized batches within current memory constraints rather than processing requests sequentially. This improves throughput and resource utilization compared to FCFS systems, especially for heterogeneous request sizes.

Our system also includes a~\textit{zero-copy mechanism} in our KV-cache memory pool, where KV-caches are managed through pointer operations rather than data movement. This design eliminates redundant memory copies between compression stages, significantly reducing memory bandwidth overhead and I/O data transfer latency. By maintaining direct references to cache locations, our memory pool achieves efficient cache state transitions without the overhead of physical data relocation.

\subsection{Evaluation Setup}
We conduct extensive experiments to evaluate the effectiveness of our system in real-world serving scenarios. All experiments are conducted on a single NVIDIA H100 GPU with 80GB HBM3 memory. Our evaluation focuses on system performance under concurrent multimodal requests while maintaining model accuracy.

\paragraph{Model and datasets.}
We evaluate our system using LLaVA-1.5-7B~\cite{liu2023llava}, a state-of-the-art multimodal large language model. For comprehensive evaluation, we utilize two distinct datasets: GQA ~\cite{hudson2019gqa}and MileBench~\cite{dingjie2024milebench}. GQA is a large-scale visual question answering dataset that features complex visual reasoning tasks. MileBench is a specialized benchmark for testing multimodal long-context capabilities, containing an average of 15.2 images and 422.3 words per sample across 6,440 multimodal instances.

To focus on system performance under realistic serving conditions, we specifically utilize the image components from both datasets while employing simple descriptive text prompts. This configuration allows us to stress-test the system's concurrent processing capabilities while maintaining controlled experimental conditions.

\paragraph{Baseline Methods.}
We compare our approach against three state-of-the-art KV-cache compression methods: SnapKV~\cite{li2024snapkv}, Knorm~\cite{devoto2024simpleeffective}, StreamingLLM~\cite{xiao2023efficient}and Nvidia KV-cache compress~\cite{kvpress2024}. For serving system comparisons, we implement multiple baseline scheduling strategies: a vanilla First-Come-First-Serve (FCFS) system and several static batch scheduling systems with different fixed batch sizes. These baselines represent current standard practices in production serving systems. We evaluate our dynamic online scheduler against these fixed scheduling approaches to demonstrate the advantages of adaptive resource management and batching strategies.

\subsection{Evaluation Metrics}
We evaluate our system through two primary dimensions: compression quality and system performance metrics. To assess compression quality, we employ ROUGE and BLEU scores as our primary metrics. ROUGE evaluates the quality of generated text by measuring recall-based overlap between the system's output and reference full kv-cache responses, while BLEU provides a complementary precision-based assessment by comparing generated responses against full kv-cache references.

For system performance evaluation, we focus on three key metrics that capture different aspects of serving efficiency. TTFT measures the initial response latency, representing the time between request submission and the generation of the first output token. This metric is crucial for assessing system responsiveness. TPOT captures the average time required to generate each subsequent token, providing insight into sustained generation performance. Finally, throughput measures the number of requests processed per second under various concurrent load conditions, reflecting the system's ability to handle multiple simultaneous requests efficiently.

These complementary metrics provide a comprehensive view of both model accuracy and system efficiency, enabling us to evaluate how our compression and scheduling strategies impact real-world serving scenarios. Through these measurements, we can assess both the qualitative impact of our compression techniques on model outputs and the quantitative improvements in serving performance.
\section{Experiments}

\subsection{Main Experimental Results} \label{experi}
\begin{figure*}[t]
   \centering
   \includegraphics[width=\textwidth]{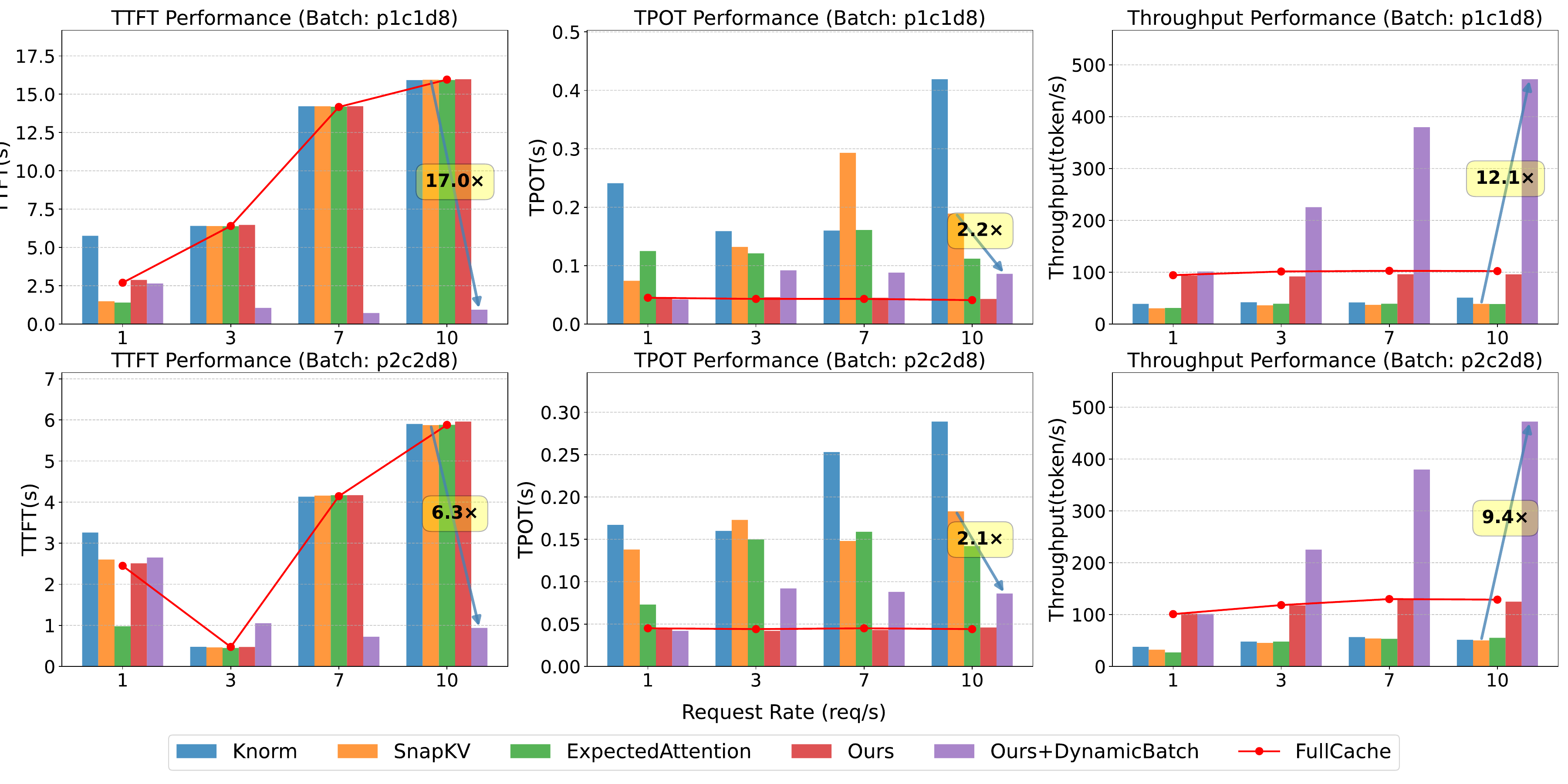}
   \caption{System performance evaluation on GQA dataset with varying request rates and static batch configurations (p1c1d8 and p2c2d8). Our method with dynamic batching achieves up to 12.1$\times$ and 9.4$\times$ throughput improvement for p1c1d8 and p2c2d8 respectively. Compared to state-of-the-art baselines, our approach reduces TTFT by up to 17.0$\times$ and TPOT by 2.2$\times$, while maintaining consistent performance advantages across different request rates and batch configurations.}
   \label{fig:milebench_results}
\end{figure*}

\begin{figure*}[t]
   \centering
   \includegraphics[width=\textwidth]{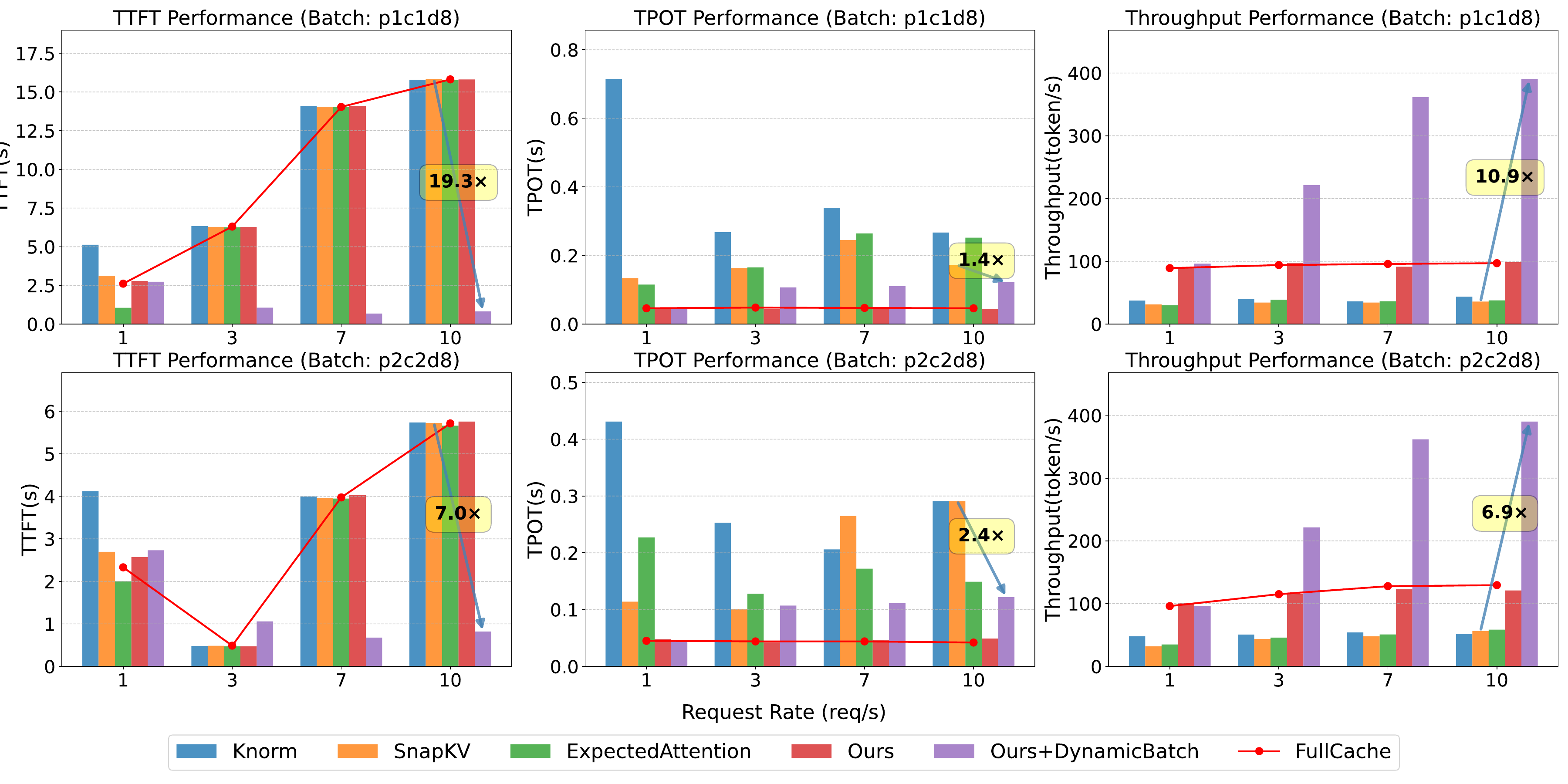}
   \caption{System performance evaluation on MileBench dataset with varying request rates and batch configurations (p1c1d8 and p2c2d8). Our method with dynamic batching demonstrates substantial improvements, achieving up to 10.9$\times$ and 6.9$\times$ throughput improvement for p1c1d8 and p2c2d8 respectively. Compared to baselines, our approach reduces TTFT by up to 19.3$\times$ and TPOT by up to 2.4$\times$, while maintaining consistent performance advantages across different request rates.}
   \label{fig:gqa_results}
\end{figure*}

We present comprehensive experimental results on GQA and MileBench datasets to evaluate our system's performance. As shown in Figure~\ref{fig:milebench_results} and Figure~\ref{fig:gqa_results}, we compare our method against several baselines including FullCache, Knorm, SnapKV, and ExpectedAttention under different request rates (1-10 req/s) and batch configurations. To thoroughly assess system behavior, we examine two static batching configurations: p1c1d8 (prefill=1, compress=1, decode=8) and p2c2d8 (prefill=2, compress=2, decode=8).

\paragraph{Latency Analysis.} \label{latency}
For TTFT, our method demonstrates substantial improvements across all request rates. As illustrated in the first row of both figures, under the p1c1d8 configuration at 10 req/s, our approach with dynamic batching reduces TTFT by up to 19.3$\times$ compared to baselines, effectively minimizing initial response latency. Even under the p2c2d8 configuration, we achieve a 7.0$\times$ reduction in TTFT, showcasing the robustness of our approach across different batch settings. This significant improvement is maintained consistently across different request rates, with particularly strong performance under high load conditions (7-10 req/s), where traditional methods struggle with resource contention and queue buildup.
\paragraph{Processing Efficiency.}
The middle row of both Figure~\ref{fig:milebench_results} and Figure~\ref{fig:gqa_results} shows TPOT results, demonstrating our system's superior processing efficiency. In both batch configurations, our method maintains consistently lower TPOT compared to baselines, indicating better token generation efficiency. For instance, under p1c1d8 configuration at 10 req/s, our approach achieves a 1.4$\times$ TPOT reduction, while under p2c2d8, we observe a 2.4$\times$ improvement. The addition of dynamic batching proves particularly effective in maintaining stable TPOT even as request rates increase, showing robust performance across varying load conditions. This stability in TPOT metrics demonstrates our system's capability to efficiently manage computational resources and maintain consistent processing performance under different workload intensities.
\paragraph{System Throughput.} \label{throughput}
The throughput measurements, shown in the bottom row of both figures, reveal the most substantial advantages of our approach. Under p1c1d8, our method with dynamic batching achieves up to 10.9$\times$ improvement in throughput compared to baselines, reaching approximately 400 tokens/s at peak load. This significant enhancement demonstrates our system's ability to effectively parallelize request processing and maximize GPU utilization. Similarly impressive results are observed in the p2c2d8 configuration, where our approach demonstrates a 6.9$\times$ throughput improvement. This dramatic enhancement in throughput efficiency is particularly notable at high request rates (7-10 req/s), where our system maintains consistent performance while other approaches show significant degradation. The sustained high throughput under increased load highlights our method's superior resource management and workload handling capabilities.
The performance patterns remain consistent between GQA and MileBench datasets, despite their different characteristics. This consistency validates the robustness of our approach and its applicability across different multimodal serving scenarios. The slight variations in absolute performance numbers between datasets can be attributed to their inherent complexity differences, but the relative improvements remain stable, confirming the generalizability of our method.

\begin{table}[t]
\centering
\caption{Strict accuracy comparison of different KV-cache compression methods on GQA. Higher scores indicate better performance.}
\label{tab:compression_accuracy}
\begin{tabular}{lccc}
\toprule
\textbf{Method} & \textbf{BLEU} & \textbf{ROUGE-1} & \textbf{ROUGE-L} \\
\midrule
SnapKV & 5.15 & 17.1 & 11.3 \\
ExpectedAttention & 7.6 & 18.1 & 12.3 \\
StreamingLLM &9.3 &15.5 & 10.9 \\
Knorm & 18.9 & 19.7 & 9.6 \\
Ours & \textbf{23.8} & \textbf{32.1} & \textbf{23.9} \\
\bottomrule
\end{tabular}
\end{table}

\subsection{System Analysis}

While our primary focus is on system performance, maintaining high model accuracy is crucial for practical deployment. Table~\ref{tab:compression_accuracy} presents a comprehensive comparison of compression quality across different methods using standard natural language generation metrics.

As shown in the results, our method significantly outperforms all other compression approaches across all evaluation metrics. Specifically, our method achieves a BLEU score of 23.8, which substantially exceeds the performance of other approaches including Knorm (18.9), StreamingLLM (9.3), ExpectedAttention (7.6), and SnapKV (5.15). This pattern of superior performance is consistently reflected in the ROUGE scores, where our method achieves notably higher scores (ROUGE-1: 32.1, ROUGE-L: 23.9) compared to the next best performing method, Knorm (ROUGE-1: 19.7) and ExpectedAttention (ROUGE-L: 12.3).

These results demonstrate that our compression approach not only preserves but enhances the quality of generated outputs compared to existing methods. The substantial improvement in accuracy metrics can be attributed to our modality-aware compression strategy and self-supervised training approach, which effectively capture and retain the most relevant features for both text and image content.
\begin{figure}[ht]
    \centering
    \includegraphics[width=0.45\textwidth]{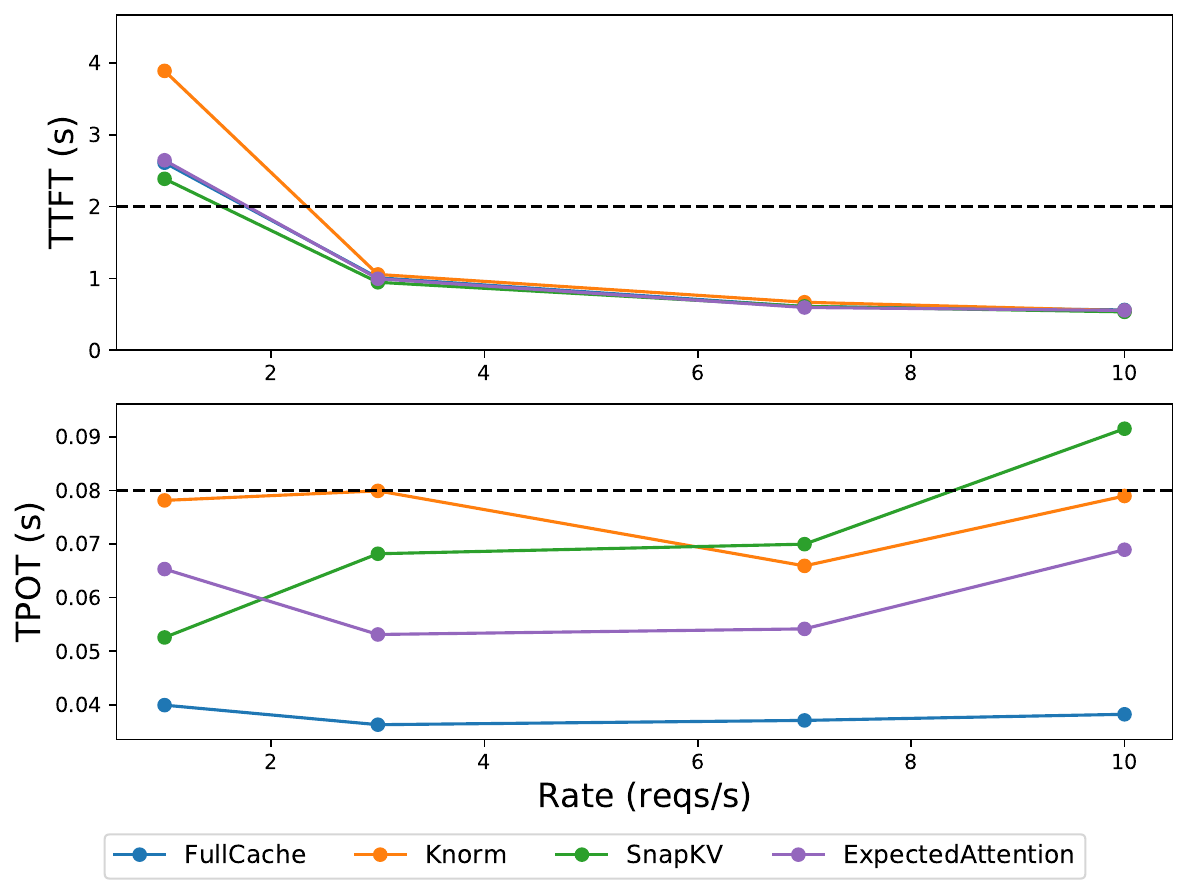}
    \caption{Performance on current compress method through \texttt{FastCache} mechanism.}
    \label{fig:online}
\end{figure}

\paragraph{Service Level Objectives.}
We evaluate the performance impact of our \texttt{FastCache} mechanism on existing KV-cache compression methods under real-time serving conditions using the LLaVA-1.5-7B model. As shown in Figure~\ref{fig:online}, after integrating with \texttt{FastCache}, all compression methods demonstrate dramatically improved performance in both TTFT and TPOT metrics. Specifically, the TTFT for all methods consistently remains under 2 seconds SLO threshold at request rates from 4 to 10 req/s, with only slight degradation at lower request rates (1-2 req/s). This represents a remarkable improvement compared to the baseline performance shown in Figure~\ref{fig:init}, where methods like Knorm peaked at 42 seconds TTFT under high load. For TPOT, while FullCache maintains the lowest values around 0.04s, all compression methods with \texttt{FastCache} achieve stable performance between 0.06-0.09s across different request rates, showing significantly reduced volatility compared to their original implementations. Most notably, even under the highest load of 10 req/s, the TTFT remains under 1 second for all methods, demonstrating that \texttt{FastCache} effectively addresses the high-latency bottlenecks of existing compression approaches while maintaining their memory efficiency benefits. These results validate that our \texttt{FastCache} mechanism successfully transforms existing KV-cache compression methods into practical solutions for real-time serving systems by eliminating their performance bottlenecks without compromising their compression capabilities.

\begin{figure}[ht]
    \centering
    \includegraphics[width=0.45\textwidth]{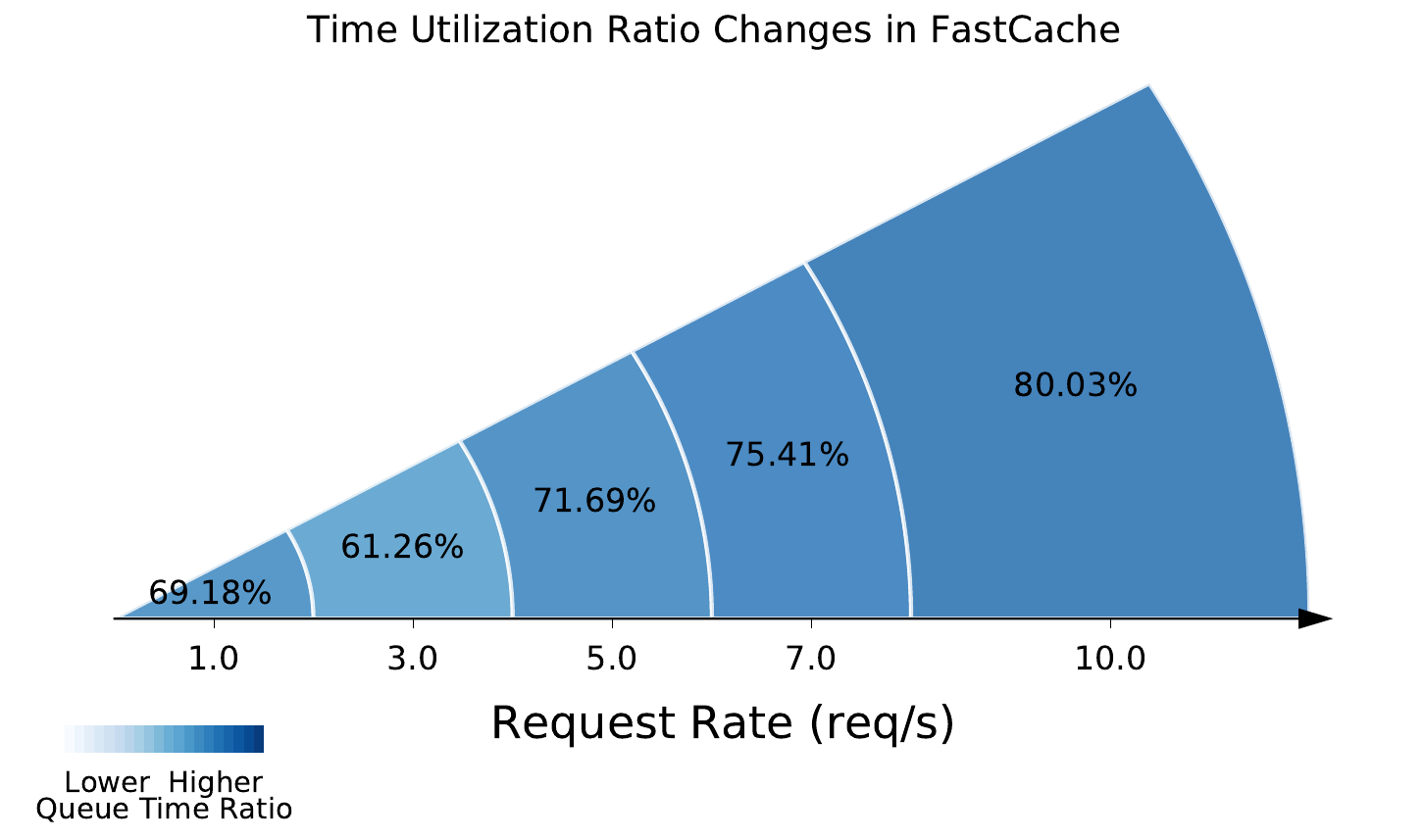}
    \caption{Time utilization ratio analysis in \texttt{FastCache} across different request rates, showing the proportion of effective processing time relative to total execution time. Higher percentages indicate more efficient resource utilization with reduced queue waiting time.}
    \label{fig:utilization}
\end{figure}
\paragraph{Queue Time Analysis.}
To evaluate the efficiency of \texttt{FastCache}'s request processing, we analyze the time utilization ratio (effective processing time versus total time) across different request rates. As shown in Figure~\ref{fig:utilization}, our system demonstrates increasingly efficient time utilization as the request rate grows. Starting from 69.18\% at 1 req/s, the utilization ratio steadily improves to 80.03\% at 10 req/s, indicating that our dynamic scheduling effectively minimizes queue waiting time even under high load. This trend suggests that \texttt{FastCache}'s resource management becomes more efficient at higher request rates, where traditional systems typically suffer from increased queuing overhead. The consistent improvement in utilization ratio (from 61.26\% at 3 req/s to 75.41\% at 7 req/s) demonstrates that our system successfully maintains low queue times relative to actual processing time, validating the effectiveness of our scheduling strategy in high-concurrency scenarios. Furthermore, comparing our FastCache system with the vanilla decoupled architecture reveals substantial improvements in resource utilization efficiency. While the decoupled prefill-decoding system (Figure~\ref{fig:time_utilization}, right) achieves a maximum utilization ratio of 55.80\% at 10 req/s, FastCache (Figure~\ref{fig:utilization}) significantly outperforms this baseline with an 80.03\% utilization ratio at the same request rate. This 24.23\% improvement in utilization ratio demonstrates that our system's sophisticated scheduling and memory management strategies effectively address the limitations of simple architectural decoupling. Notably, \texttt{FastCache} maintains consistently higher utilization ratios across all request rates, with even its lowest utilization (69.18\% at 1 req/s) exceeding the peak performance of the basic decoupled system. This comprehensive improvement validates that our framework's innovations go beyond simple architectural optimizations, effectively transforming the efficiency of LLM serving systems under compression.
\subsection{High Load and Memory Utilization Analysis}
\begin{figure}[t]
\centering
\includegraphics[width=0.3\textwidth]{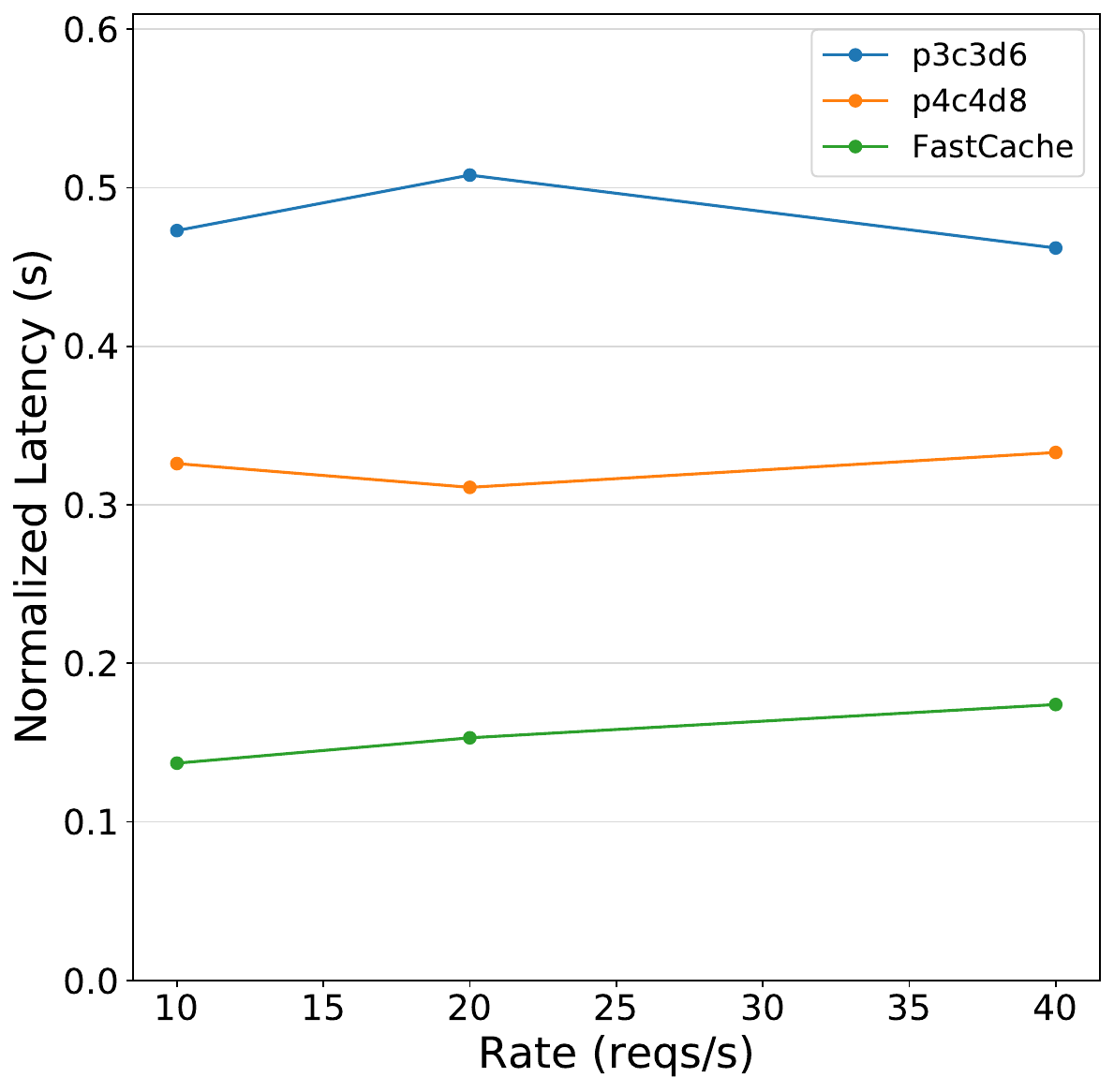}
\caption{Performance comparison of our framework against other static configurations under high-concurrency scenarios, demonstrating normalized latency across varying request rates from 10 to 40 req/s.}
\label{fig:highload}
\end{figure}

\paragraph{High Concurrency Performance Analysis.}
To further evaluate the scalability of \texttt{FastCache} under high-load scenarios, we conducted experiments with significantly increased request rates ranging from 10 to 40 req/s, comparing against two static configurations (p3c3d6 and p4c4d8). As illustrated in Figure~\ref{fig:highload}, \texttt{FastCache} demonstrates superior latency characteristics across all tested request rates. At 20 req/s, while the p3c3d6 configuration exhibits a normalized latency peak of approximately 0.51s and p4c4d8 maintains around 0.31s, \texttt{FastCache} achieves a substantially lower latency of 0.15s. This performance advantage becomes even more pronounced at higher request rates - at 40 req/s, \texttt{FastCache} maintains a stable normalized latency of 0.17s, showing only minimal degradation from its performance at lower loads. In contrast, the static configurations exhibit significantly higher latencies (0.46s for p3c3d6 and 0.33s for p4c4d8) and greater performance variability. This robust performance under extreme load conditions demonstrates that \texttt{FastCache}'s dynamic optimization strategy effectively mitigates the resource contention issues that typically plague static configurations in high-concurrency environments.

\begin{figure}[t]
\centering
\includegraphics[width=0.5\textwidth]{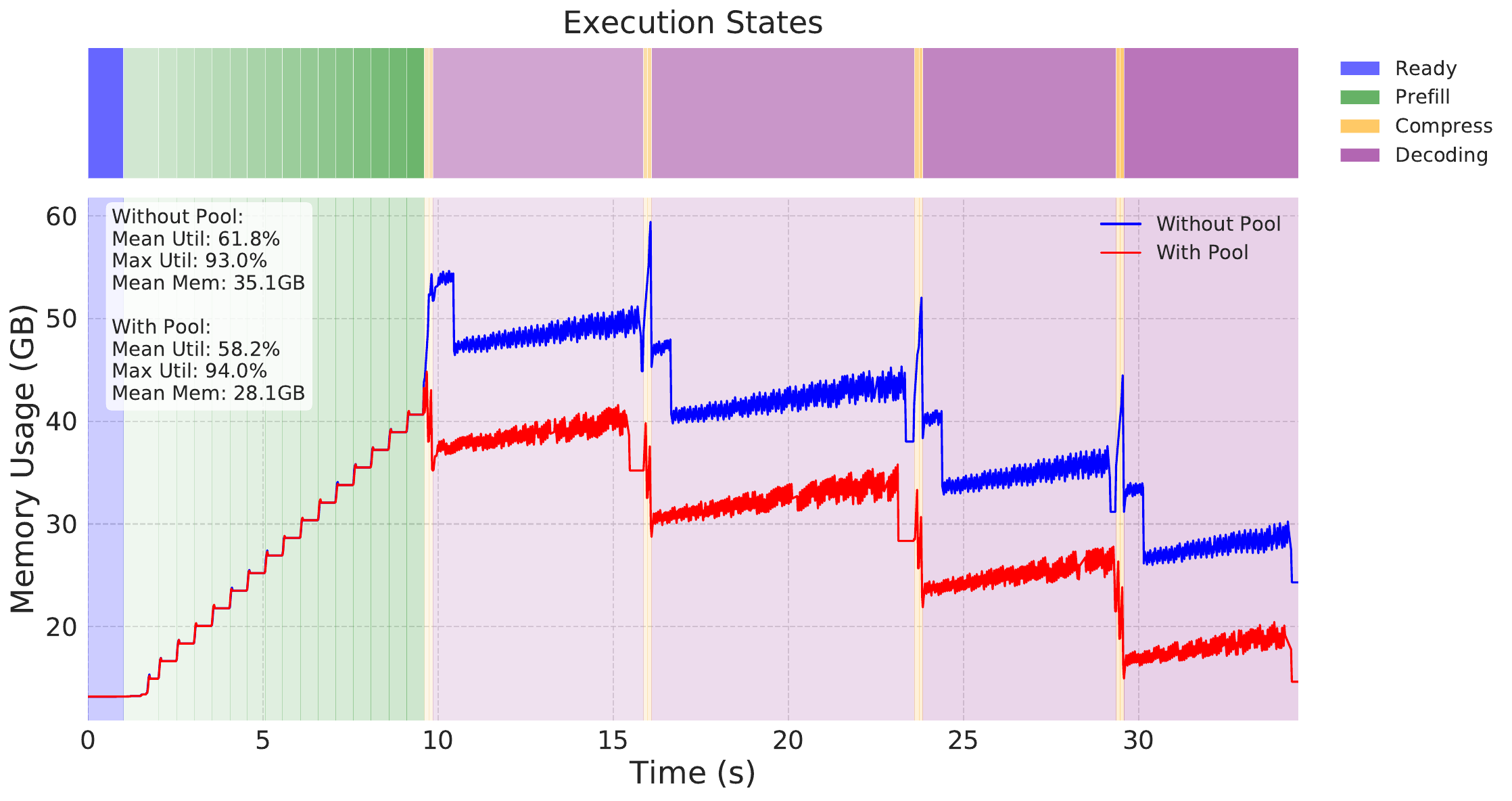}
\caption{Comparison of GPU memory usage patterns between identical configurations with and without KV-cache memory pool. Different execution states are indicated by background colors, showing significant memory savings through efficient pool-based memory management}
\label{fig:memory}
\end{figure}

\paragraph{Memory Efficiency Analysis.}
To evaluate the memory efficiency of \texttt{FastCache}'s memory pool mechanism, we conducted experiments under identical model configurations and workload settings. As shown in Figure~\ref{fig:memory}, with the same configuration, the baseline system without memory pool exhibits consistent memory fragmentation, maintaining a higher average memory consumption of 35.1GB and maximum GPU memory utilization of 93.0\%. In contrast, our memory pool mechanism demonstrates superior memory management, reducing the average memory usage to 28.1GB while achieving comparable maximum GPU utilization (94.0\%). The memory usage trajectory with pool enabled (red line) shows clear stepwise patterns and effective memory reclamation after each execution phase, particularly evident in later stages where memory consumption stabilizes around 20GB compared to the baseline's 30GB. This 20\% reduction in average memory usage without compromising computational efficiency validates that our memory pool design effectively eliminates memory fragmentation issues in continuous serving scenarios.
\subsection{Ablation Study}
We conduct ablation experiments to evaluate the individual contributions of our two key system components: online ( dynamic) batching and memory pool management.
\paragraph{Impact of Dynamic Batching.}
Comparing the performance between static and dynamic batching configurations (shown in Figure~\ref{fig:milebench_results} and Figure~\ref{fig:gqa_results}, "Ours" vs "Ours+DynamicBatch"), we observe that dynamic batching provides substantial improvements. Under certain load scenarios (10 req/s), it enables up to 19.3$\times$ TTFT reduction while improving throughput by 10.9$\times$, demonstrating its effectiveness in optimizing request processing efficiency.
\paragraph{Memory Pool Management.}
The memory pool mechanism (Figure~\ref{fig:memory}) reduces average memory consumption from 35.1GB to 28.1GB while maintaining comparable GPU utilization (93.0\% vs 94.0\%). The stepwise memory usage pattern demonstrates effective memory reclamation across different execution states, particularly evident in the stabilized memory consumption around 20GB during later execution phases.
These results validate that both components are essential for achieving optimal system performance, with dynamic batching primarily improving processing efficiency while the memory pool optimizes resource utilization.

\section{Related Works} \label{related}

\paragraph{LLM Inference.} Modern LLMs predict the next token given an input sequence by computing hidden representations for each token. While an LLM can process variable-length inputs in parallel, its computational workload increases superlinearly with the number of tokens. This computation demands substantial I/O to move LLM weights and intermediate states between GPU's HBM and SRAM~\cite{dao2022flashattention,dao2023flashattention}.
During the inference process, KV-cache emerges as a critical component across both prefill and decode phases. In prefill, the model processes multiple tokens concurrently, generating key-value pairs that serve as intermediate states. During decode, despite handling just one new token per step, the model requires access to both model weights and these cached states~\cite{nawrot2024dynamic}. The importance of KV-cache lies in its dual role: it eliminates the need for expensive recomputation of previous tokens' representations while creating a significant memory footprint that grows linearly with sequence length.
Since LLM inference engines typically colocate prefill and decode phases on GPUs due to their shared resource requirements, efficient KV-cache management becomes crucial for balancing computational efficiency against memory constraints. This makes it a central challenge in LLM serving systems, particularly in high-throughput scenarios where multiple requests compete for limited GPU memory.

\paragraph{KV-Cache Management Methods.}
Current methods for KV-cache compression can be broadly categorized into three main approaches. First, \textbf{token-level optimization}~\cite{ge2023model,li2024snapkv,sheng2023flexgen}involves selectively storing key tokens, dynamically allocating memory budgets, merging similar key-value pairs~\cite{wan2024look, wang2024model, kim2023compressed}, reducing precision through quantization, and applying low-rank decomposition to minimize redundancy and memory usage. Second, \textbf{model-level optimization}~\cite{shazeer2019fast,ainslie2023gqa,chen2024optimised,chinnakonduru2024weighted} focuses on improving model architecture by grouping and sharing attention mechanisms, introducing new architectural designs, or integrating non-Transformer architectures to enhance the efficiency of KV cache reuse. Third, \textbf{system-level optimization}~\cite{kwon2023efficient,zheng2023efficiently,xu2024vtensor} employs advanced memory management and scheduling strategies, such as virtual memory techniques, intelligent prefix sharing~\cite{gim2024prompt, juravsky2024hydragen}, and hierarchical scheduling~\cite{gao2024cost, zhao2024alisa}, to optimize resource utilization across different computing environments.

\paragraph{LLM Serving Systems.}
Recent LLM serving systems focus on resource and memory optimization. vLLM~\cite{kwon2023efficient} uses paged attention for GPU memory efficiency, FlexGen~\cite{sheng2023flexgen} leverages hybrid CPU-GPU execution, and DistServe~\cite{zhong2024distserve} introduces specialized Prefill/Decode nodes. While these advances improve resource utilization, they assume uncompressed KV-caches and cannot handle the overhead from KV-cache compression.
Current KV-cache compression methods rely heavily on attention-based approaches, requiring frequent access to attention scores and KV-pairs~\cite{dao2022flashattention}. This creates memory bandwidth bottlenecks as compression operations compete with inference tasks for memory access. The memory-bound nature of these techniques, when integrated into serving pipelines, severely impacts performance.
The addition of KV-cache compression fundamentally changes resource utilization patterns in serving systems. Traditional serving architectures, optimized for prefill-decode workflows, become inefficient when compression is introduced. This reveals a crucial gap: the need for end-to-end architectures that efficiently handle compression overhead while maintaining optimal resource utilization.

Our work addresses this challenge by introducing a comprehensive KV-compress framework that integrates adaptive compression strategies with dynamic scheduling mechanisms, specifically designed to optimize performance in the presence of KV-cache memory pool.
\section{Future Work}
While \texttt{FastCache} demonstrates significant improvements in KV-cache compression efficiency, several promising directions remain for future exploration. First, we plan to extend our memory pool mechanism to support heterogeneous hardware configurations, particularly focusing on multi-GPU scenarios where memory management becomes more complex due to cross-device communication and synchronization. Our KV-cache memory pool could be further enhanced through prefetching mechanisms to predict and preload frequently accessed cache entries, enabling support for higher request loads. Additionally, introducing CPU-GPU hybrid memory management~\cite{jiang2020unified,zhang2020finestream} could leverage CPU memory as an additional tier for asynchronous cache management, significantly improving system throughput in high-concurrency scenarios.
Furthermore, investigating adaptive compression strategies that dynamically adjust compression ratios based on real-time workload characteristics and resource availability could further optimize memory efficiency. Exploring the integration of  \texttt{FastCache} with emerging model serving frameworks and different model architectures (e.g., Mixture-of-Experts~\cite{chen2022towards,zhou2022mixture}) could broaden its applicability, while developing automated tuning mechanisms for pool configuration parameters based on historical serving patterns could reduce manual optimization efforts and improve out-of-the-box performance.

\section{Conclusion} \label{conclu}
In this paper, we present \texttt{FastCache}, a novel serving framework that addresses the critical challenges of KV-cache compression in MLLM inference. Our system introduces two key innovations: a dynamic batching strategy that significantly improves request processing efficiency, and an efficient memory pool mechanism that effectively manages GPU memory resources. Through extensive experiments on multimodal datasets, \texttt{FastCache} demonstrates substantial performance improvements over existing approaches, achieving up to 19.3$\times$ reduction in TTFT and 12.1$\times$ improvement in throughput while maintaining stable TPOT metrics. The memory pool mechanism further reduces average memory consumption by 20\% without compromising computational efficiency. Our system remains robust under high-concurrency scenarios, maintaining stable performance even at 40 requests per second. These results validate that \texttt{FastCache} successfully transforms existing KV-cache compression methods into practical solutions for real-time serving systems.

{\footnotesize \bibliographystyle{acm}
\bibliography{sample}}

\begin{thebibliography}{10}

\bibitem{adnan2024keyformer}
{\sc Adnan, M., Arunkumar, A., Jain, G., Nair, P., Soloveychik, I., and Kamath, P.}
\newblock Keyformer: Kv cache reduction through key tokens selection for efficient generative inference.
\newblock {\em Proceedings of Machine Learning and Systems 6\/} (2024), 114--127.

\bibitem{agrawal2024taming}
{\sc Agrawal, A., Kedia, N., Panwar, A., Mohan, J., Kwatra, N., Gulavani, B., Tumanov, A., and Ramjee, R.}
\newblock Taming $\{$Throughput-Latency$\}$ tradeoff in $\{$LLM$\}$ inference with $\{$Sarathi-Serve$\}$.
\newblock In {\em 18th USENIX Symposium on Operating Systems Design and Implementation (OSDI 24)\/} (2024), pp.~117--134.

\bibitem{ainslie2023gqa}
{\sc Ainslie, J., Lee-Thorp, J., de~Jong, M., Zemlyanskiy, Y., Lebr{\'o}n, F., and Sanghai, S.}
\newblock Gqa: Training generalized multi-query transformer models from multi-head checkpoints.
\newblock {\em arXiv preprint arXiv:2305.13245\/} (2023).

\bibitem{chen2024optimised}
{\sc Chen, Y., Zhang, C., Gao, X., Mullins, R.~D., Constantinides, G.~A., and Zhao, Y.}
\newblock Optimised grouped-query attention mechanism for transformers.
\newblock {\em arXiv preprint arXiv:2406.14963\/} (2024).

\bibitem{chen2022towards}
{\sc Chen, Z., Deng, Y., Wu, Y., Gu, Q., and Li, Y.}
\newblock Towards understanding the mixture-of-experts layer in deep learning.
\newblock {\em Advances in neural information processing systems 35\/} (2022), 23049--23062.

\bibitem{chinnakonduru2024weighted}
{\sc Chinnakonduru, S.~S., and Mohapatra, A.}
\newblock Weighted grouped query attention in transformers.
\newblock {\em arXiv preprint arXiv:2407.10855\/} (2024).

\bibitem{cho2024starnuma}
{\sc Cho, A., and Daglis, A.}
\newblock Starnuma: Mitigating numa challenges with memory pooling.
\newblock In {\em 2024 57th IEEE/ACM International Symposium on Microarchitecture (MICRO)\/} (2024), IEEE, pp.~997--1012.

\bibitem{crankshaw2017clipper}
{\sc Crankshaw, D., Wang, X., Zhou, G., Franklin, M.~J., Gonzalez, J.~E., and Stoica, I.}
\newblock Clipper: A $\{$Low-Latency$\}$ online prediction serving system.
\newblock In {\em 14th USENIX Symposium on Networked Systems Design and Implementation (NSDI 17)\/} (2017), pp.~613--627.

\bibitem{dai2024apparate}
{\sc Dai, Y., Pan, R., Iyer, A., Li, K., and Netravali, R.}
\newblock Apparate: Rethinking early exits to tame latency-throughput tensions in ml serving.
\newblock In {\em Proceedings of the ACM SIGOPS 30th Symposium on Operating Systems Principles\/} (2024), pp.~607--623.

\bibitem{dao2023flashattention}
{\sc Dao, T.}
\newblock Flashattention-2: Faster attention with better parallelism and work partitioning.
\newblock {\em arXiv preprint arXiv:2307.08691\/} (2023).

\bibitem{dao2022flashattention}
{\sc Dao, T., Fu, D., Ermon, S., Rudra, A., and R{\'e}, C.}
\newblock Flashattention: Fast and memory-efficient exact attention with io-awareness.
\newblock {\em Advances in Neural Information Processing Systems 35\/} (2022), 16344--16359.

\bibitem{devoto2024simpleeffective}
{\sc Devoto, A., Zhao, Y., Scardapane, S., and Minervini, P.}
\newblock A simple and effective {$L_2$} norm-based strategy for {KV} cache compression.
\newblock In {\em Proceedings of the 2024 Conference on Empirical Methods in Natural Language Processing (EMNLP)\/} (2024).

\bibitem{dingjie2024milebench}
{\sc Dingjie, S., Chen, S., Chen, G.~H., Yu, F., Wan, X., and Wang, B.}
\newblock Milebench: Benchmarking mllms in long context.
\newblock In {\em First Conference on Language Modeling\/} (2024).

\bibitem{fu2024serverlessllm}
{\sc Fu, Y., Xue, L., Huang, Y., Brabete, A.-O., Ustiugov, D., Patel, Y., and Mai, L.}
\newblock Serverlessllm: Low-latency serverless inference for large language models.
\newblock In {\em 18th USENIX Symposium on Operating Systems Design and Implementation\/} (2024), USENIX Association, pp.~135--153.

\bibitem{gao2024cost}
{\sc Gao, B., He, Z., Sharma, P., Kang, Q., Jevdjic, D., Deng, J., Yang, X., Yu, Z., and Zuo, P.}
\newblock $\{$Cost-Efficient$\}$ large language model serving for multi-turn conversations with $\{$CachedAttention$\}$.
\newblock In {\em 2024 USENIX Annual Technical Conference (USENIX ATC 24)\/} (2024), pp.~111--126.

\bibitem{ge2023model}
{\sc Ge, S., Zhang, Y., Liu, L., Zhang, M., Han, J., and Gao, J.}
\newblock Model tells you what to discard: Adaptive kv cache compression for llms.
\newblock {\em arXiv preprint arXiv:2310.01801\/} (2023).

\bibitem{gim2024prompt}
{\sc Gim, I., Chen, G., Lee, S.-s., Sarda, N., Khandelwal, A., and Zhong, L.}
\newblock Prompt cache: Modular attention reuse for low-latency inference.
\newblock {\em Proceedings of Machine Learning and Systems 6\/} (2024), 325--338.

\bibitem{he2022masked}
{\sc He, K., Chen, X., Xie, S., Li, Y., Doll{\'a}r, P., and Girshick, R.}
\newblock Masked autoencoders are scalable vision learners.
\newblock In {\em Proceedings of the IEEE/CVF conference on computer vision and pattern recognition\/} (2022), pp.~16000--16009.

\bibitem{hudson2019gqa}
{\sc Hudson, D.~A., and Manning, C.~D.}
\newblock Gqa: A new dataset for real-world visual reasoning and compositional question answering.
\newblock In {\em Proceedings of the IEEE/CVF conference on computer vision and pattern recognition\/} (2019), pp.~6700--6709.

\bibitem{jiang2020unified}
{\sc Jiang, Y., Zhu, Y., Lan, C., Yi, B., Cui, Y., and Guo, C.}
\newblock A unified architecture for accelerating distributed $\{$DNN$\}$ training in heterogeneous $\{$GPU/CPU$\}$ clusters.
\newblock In {\em 14th USENIX Symposium on Operating Systems Design and Implementation (OSDI 20)\/} (2020), pp.~463--479.

\bibitem{juravsky2024hydragen}
{\sc Juravsky, J., Brown, B., Ehrlich, R., Fu, D.~Y., R{\'e}, C., and Mirhoseini, A.}
\newblock Hydragen: High-throughput llm inference with shared prefixes.
\newblock {\em arXiv preprint arXiv:2402.05099\/} (2024).

\bibitem{kim2023compressed}
{\sc Kim, J.-H., Yeom, J., Yun, S., and Song, H.~O.}
\newblock Compressed context memory for online language model interaction.
\newblock {\em arXiv preprint arXiv:2312.03414\/} (2023).

\bibitem{kwon2023efficient}
{\sc Kwon, W., Li, Z., Zhuang, S., Sheng, Y., Zheng, L., Yu, C.~H., Gonzalez, J., Zhang, H., and Stoica, I.}
\newblock Efficient memory management for large language model serving with pagedattention.
\newblock In {\em Proceedings of the 29th Symposium on Operating Systems Principles\/} (2023), pp.~611--626.

\bibitem{lee2024infinigen}
{\sc Lee, W., Lee, J., Seo, J., and Sim, J.}
\newblock $\{$InfiniGen$\}$: Efficient generative inference of large language models with dynamic $\{$KV$\}$ cache management.
\newblock In {\em 18th USENIX Symposium on Operating Systems Design and Implementation (OSDI 24)\/} (2024), pp.~155--172.

\bibitem{li2023pond}
{\sc Li, H., Berger, D.~S., Hsu, L., Ernst, D., Zardoshti, P., Novakovic, S., Shah, M., Rajadnya, S., Lee, S., Agarwal, I., et~al.}
\newblock Pond: Cxl-based memory pooling systems for cloud platforms.
\newblock In {\em Proceedings of the 28th ACM International Conference on Architectural Support for Programming Languages and Operating Systems, Volume 2\/} (2023), pp.~574--587.

\bibitem{li2024snapkv}
{\sc Li, Y., Huang, Y., Yang, B., Venkitesh, B., Locatelli, A., Ye, H., Cai, T., Lewis, P., and Chen, D.}
\newblock Snapkv: Llm knows what you are looking for before generation.
\newblock {\em arXiv preprint arXiv:2404.14469\/} (2024).

\bibitem{liu2023llava}
{\sc Liu, H., Li, C., Wu, Q., and Lee, Y.~J.}
\newblock Visual instruction tuning, 2023.

\bibitem{liu2024cachegen}
{\sc Liu, Y., Li, H., Cheng, Y., Ray, S., Huang, Y., Zhang, Q., Du, K., Yao, J., Lu, S., Ananthanarayanan, G., et~al.}
\newblock Cachegen: Kv cache compression and streaming for fast large language model serving.
\newblock In {\em Proceedings of the ACM SIGCOMM 2024 Conference\/} (2024), pp.~38--56.

\bibitem{liu2024scissorhands}
{\sc Liu, Z., Desai, A., Liao, F., Wang, W., Xie, V., Xu, Z., Kyrillidis, A., and Shrivastava, A.}
\newblock Scissorhands: Exploiting the persistence of importance hypothesis for llm kv cache compression at test time.
\newblock {\em Advances in Neural Information Processing Systems 36\/} (2024).

\bibitem{liu2024kivi}
{\sc Liu, Z., Yuan, J., Jin, H., Zhong, S., Xu, Z., Braverman, V., Chen, B., and Hu, X.}
\newblock Kivi: A tuning-free asymmetric 2bit quantization for kv cache.
\newblock In {\em Proceedings of the 41st International Conference on Machine Learning\/} (2024).

\bibitem{mitropoulou2016lynx}
{\sc Mitropoulou, K., Porpodas, V., Zhang, X., and Jones, T.~M.}
\newblock Lynx: Using os and hardware support for fast fine-grained inter-core communication.
\newblock In {\em Proceedings of the 2016 International Conference on Supercomputing\/} (2016), pp.~1--12.

\bibitem{narayanan2021memory}
{\sc Narayanan, D., Phanishayee, A., Shi, K., Chen, X., and Zaharia, M.}
\newblock Memory-efficient pipeline-parallel dnn training.
\newblock In {\em International Conference on Machine Learning\/} (2021), PMLR, pp.~7937--7947.

\bibitem{nawrot2024dynamic}
{\sc Nawrot, P., {\L}a{\'n}cucki, A., Chochowski, M., Tarjan, D., and Ponti, E.~M.}
\newblock Dynamic memory compression: Retrofitting llms for accelerated inference.
\newblock {\em arXiv preprint arXiv:2403.09636\/} (2024).

\bibitem{kvpress2024}
{\sc NVIDIA}.
\newblock Kvpress: An nvidia hardware-accelerated framework for key-value cache compression.
\newblock \url{https://github.com/NVIDIA/kvpress}, 2024.
\newblock Accessed: 2024-03-30.

\bibitem{openai2023gpt}
{\sc OpenAI, R.}
\newblock Gpt-4 technical report. arxiv 2303.08774.
\newblock {\em View in Article 2}, 5 (2023).

\bibitem{patel2024splitwise}
{\sc Patel, P., Choukse, E., Zhang, C., Shah, A., Goiri, {\'I}., Maleki, S., and Bianchini, R.}
\newblock Splitwise: Efficient generative llm inference using phase splitting.
\newblock In {\em 2024 ACM/IEEE 51st Annual International Symposium on Computer Architecture (ISCA)\/} (2024), IEEE, pp.~118--132.

\bibitem{pope2023efficiently}
{\sc Pope, R., Douglas, S., Chowdhery, A., Devlin, J., Bradbury, J., Heek, J., Xiao, K., Agrawal, S., and Dean, J.}
\newblock Efficiently scaling transformer inference.
\newblock {\em Proceedings of Machine Learning and Systems 5\/} (2023), 606--624.

\bibitem{qin2024mooncake}
{\sc Qin, R., Li, Z., He, W., Zhang, M., Wu, Y., Zheng, W., and Xu, X.}
\newblock Mooncake: A kvcache-centric disaggregated architecture for llm serving.
\newblock {\em arXiv preprint arXiv:2407.00079\/} (2024).

\bibitem{shazeer2019fast}
{\sc Shazeer, N.}
\newblock Fast transformer decoding: One write-head is all you need.
\newblock {\em arXiv preprint arXiv:1911.02150\/} (2019).

\bibitem{sheng2023flexgen}
{\sc Sheng, Y., Zheng, L., Yuan, B., Li, Z., Ryabinin, M., Chen, B., Liang, P., R{\'e}, C., Stoica, I., and Zhang, C.}
\newblock Flexgen: High-throughput generative inference of large language models with a single gpu.
\newblock In {\em International Conference on Machine Learning\/} (2023), PMLR, pp.~31094--31116.

\bibitem{shoeybi2019megatron}
{\sc Shoeybi, M., Patwary, M., Puri, R., LeGresley, P., Casper, J., and Catanzaro, B.}
\newblock Megatron-lm: Training multi-billion parameter language models using model parallelism.
\newblock {\em arXiv preprint arXiv:1909.08053\/} (2019).

\bibitem{wan2024look}
{\sc Wan, Z., Wu, Z., Liu, C., Huang, J., Zhu, Z., Jin, P., Wang, L., and Yuan, L.}
\newblock Look-m: Look-once optimization in kv cache for efficient multimodal long-context inference.
\newblock {\em Findings of the Association for Computational Linguistics\/} (2024).

\bibitem{wang2024model}
{\sc Wang, Z., Jin, B., Yu, Z., and Zhang, M.}
\newblock Model tells you where to merge: Adaptive kv cache merging for llms on long-context tasks.
\newblock {\em arXiv preprint arXiv:2407.08454\/} (2024).

\bibitem{wu2024loongserve}
{\sc Wu, B., Liu, S., Zhong, Y., Sun, P., Liu, X., and Jin, X.}
\newblock Loongserve: Efficiently serving long-context large language models with elastic sequence parallelism.
\newblock In {\em Proceedings of the ACM SIGOPS 30th Symposium on Operating Systems Principles\/} (2024), pp.~640--654.

\bibitem{wu2024dlora}
{\sc Wu, B., Zhu, R., Zhang, Z., Sun, P., Liu, X., and Jin, X.}
\newblock $\{$dLoRA$\}$: Dynamically orchestrating requests and adapters for $\{$LoRA$\}$$\{$LLM$\}$ serving.
\newblock In {\em 18th USENIX Symposium on Operating Systems Design and Implementation (OSDI 24)\/} (2024), pp.~911--927.

\bibitem{xiao2023efficient}
{\sc Xiao, G., Tian, Y., Chen, B., Han, S., and Lewis, M.}
\newblock Efficient streaming language models with attention sinks.
\newblock {\em arXiv preprint arXiv:2309.17453\/} (2023).

\bibitem{xu2024vtensor}
{\sc Xu, J., Zhang, R., Guo, C., Hu, W., Liu, Z., Wu, F., Feng, Y., Sun, S., Shao, C., Guo, Y., et~al.}
\newblock vtensor: Flexible virtual tensor management for efficient llm serving.
\newblock {\em arXiv preprint arXiv:2407.15309\/} (2024).

\bibitem{zhang2020finestream}
{\sc Zhang, F., Yang, L., Zhang, S., He, B., Lu, W., and Du, X.}
\newblock $\{$FineStream$\}$:$\{$Fine-Grained$\}$$\{$Window-Based$\}$ stream processing on $\{$CPU-GPU$\}$ integrated architectures.
\newblock In {\em 2020 USENIX Annual Technical Conference (USENIX ATC 20)\/} (2020), pp.~633--647.

\bibitem{zhang2024pyramidkv}
{\sc Zhang, Y., Gao, B., Liu, T., Lu, K., Xiong, W., Dong, Y., Chang, B., Hu, J., Xiao, W., et~al.}
\newblock Pyramidkv: Dynamic kv cache compression based on pyramidal information funneling.
\newblock {\em arXiv preprint arXiv:2406.02069\/} (2024).

\bibitem{zhang2023h2o}
{\sc Zhang, Z., Sheng, Y., Zhou, T., Chen, T., Zheng, L., Cai, R., Song, Z., Tian, Y., R{\'e}, C., Barrett, C., et~al.}
\newblock H2o: Heavy-hitter oracle for efficient generative inference of large language models.
\newblock {\em Advances in Neural Information Processing Systems 36\/} (2023), 34661--34710.

\bibitem{zhao2024alisa}
{\sc Zhao, Y., Wu, D., and Wang, J.}
\newblock Alisa: Accelerating large language model inference via sparsity-aware kv caching.
\newblock {\em arXiv preprint arXiv:2403.17312\/} (2024).

\bibitem{zheng2023efficiently}
{\sc Zheng, L., Yin, L., Xie, Z., Huang, J., Sun, C., Yu, C., Cao, S., Kozyrakis, C., Stoica, I., Gonzalez, J.~E., et~al.}
\newblock Efficiently programming large language models using sglang.

\bibitem{zheng2024sglang}
{\sc Zheng, L., Yin, L., Xie, Z., Sun, C., Huang, J., Yu, C.~H., Cao, S., Kozyrakis, C., Stoica, I., Gonzalez, J.~E., et~al.}
\newblock Sglang: Efficient execution of structured language model programs.
\newblock {\em arXiv preprint arXiv:2312.07104\/} (2024).

\bibitem{zhong2024distserve}
{\sc Zhong, Y., Liu, S., Chen, J., Hu, J., Zhu, Y., Liu, X., Jin, X., and Zhang, H.}
\newblock $\{$DistServe$\}$: Disaggregating prefill and decoding for goodput-optimized large language model serving.
\newblock In {\em 18th USENIX Symposium on Operating Systems Design and Implementation (OSDI 24)\/} (2024), pp.~193--210.

\bibitem{zhou2022mixture}
{\sc Zhou, Y., Lei, T., Liu, H., Du, N., Huang, Y., Zhao, V., Dai, A.~M., Le, Q.~V., Laudon, J., et~al.}
\newblock Mixture-of-experts with expert choice routing.
\newblock {\em Advances in Neural Information Processing Systems 35\/} (2022), 7103--7114.

\end{thebibliography}


\end{document}